\documentclass[conference]{IEEEtran}
\IEEEoverridecommandlockouts
\usepackage{cite}
\usepackage{amsmath,amssymb,amsfonts}

\DeclareMathOperator*{\argmax}{arg\,max}
\usepackage{algorithmic}
\usepackage{graphicx}
\usepackage{textcomp}
\usepackage{xcolor}
\usepackage{algorithm2e}
\usepackage{hyperref}
\usepackage{subfigure}
\def\BibTeX{{\rm B\kern-.05em{\sc i\kern-.025em b}\kern-.08em
T\kern-.1667em\lower.7ex\hbox{E}\kern-.125emX}}

\newcommand{\MyColor}{black} 

\begin{document}
\makeatletter
\newcommand{\linebreakand}{%
  \end{@IEEEauthorhalign}
  \hfill\mbox{}\par
  \mbox{}\hfill\begin{@IEEEauthorhalign}
}
\makeatother
\author{
    \IEEEauthorblockN{Ami Berger\IEEEauthorrefmark{1}, Vladimir Tourbabin\IEEEauthorrefmark{2}, Jacob Donley\IEEEauthorrefmark{2}, Zamir Ben-Hur\IEEEauthorrefmark{2}, Boaz Rafaely\IEEEauthorrefmark{1}}
    \IEEEauthorblockA{\IEEEauthorrefmark{1}School of Electrical and Computer Engineering, Ben-Gurion University of the Negev, Beer-Sheva 84105, Israel}
    \IEEEauthorblockA{\IEEEauthorrefmark{2}Reality Labs Research @ Meta, Menlo Park, CA 94025, USA}
}

\title{\textcolor{\MyColor}{Performance and Robustness of Signal-Dependent vs. Signal-Independent Binaural Signal Matching with Wearable Microphone Arrays}}
\maketitle
\thispagestyle{plain}
\pagestyle{plain}

\begin{abstract}
The increasing popularity of spatial audio in applications such as teleconferencing, entertainment, and virtual reality has led to the recent developments of binaural reproduction methods. However, only a few of these methods are well-suited for wearable and mobile arrays, which typically consist of a small number of microphones.
One such method is binaural signal matching (BSM), which has been shown to produce high-quality binaural signals for wearable arrays. However, BSM may be suboptimal in cases of high direct-to-reverberant ratio (DRR) as it is based on the diffuse sound field assumption.
To overcome this limitation, previous studies incorporated sound-field models other than diffuse. \textcolor{\MyColor}{However, performance may be sensitive to signal estimation errors.} 
This paper \textcolor{\MyColor}{aims to provide a systematic and comprehensive analysis of signal-dependent vs. signal-independent BSM, so that the benefits and limitations of the methods become clearer. Two signal-dependent} BSM-based methods designed for high DRR scenarios that incorporate a sound field model composed of direct and reverberant components are investigated mathematically, using simulations, and finally validated by a listening test, \textcolor{\MyColor}{and compared to the signal-independent BSM}. The results show that \textcolor{\MyColor}{signal-dependent BSM} can significantly improve performance, in particular in the direction of the source, while presenting only a negligible degradation in other directions.  Furthermore, when source direction estimation is inaccurate, performance of \textcolor{\MyColor}{of the signal-dependent BSM} degrade to equal that of the \textcolor{\MyColor}{signal-independent} BSM, presenting a desired robustness quality.

\end{abstract}


\section{Introduction}
The binaural reproduction of acoustic scenes captured by microphone arrays is gaining popularity in diverse applications, such as teleconferencing and virtual and augmented reality, as evident by recent publications \cite{ref1,ref2,ref3}. 
A common technique for binaural reproduction involves convolving high-order Ambisonics (HOA) signals with the head-related transfer function (HRTF) \cite{ref4}. This method is relatively precise, especially for sufficiently high spherical harmonics (SH) orders, and it can provide a realistic immersion experience through the integration of head tracking \cite{ref2}. However, the primary constraint of this method is the requirement for an extensive number of microphones when the audio signal is captured using a microphone array, as well as the need for a spherical array geometry, which limits the practical application of this method \cite{ref2}.

To address the limitations of binaural signal reproduction with spherical array geometries, the beamforming-based binaural reproduction (BFBR) technique was proposed \cite{ref5,ref6,ref7}. In this method, microphone signals are filtered using a set of beamformers, followed by filtering with the head-related transfer functions (HRTFs), finally added to produce the binaural signals. A theoretical framework for determining the design parameters of BFBR, such as the number of beams and their look direction for both planar and spherical arrays, was presented in \cite{ref8}. However, the guidelines for more general array geometries were limited and did not guarantee accurate binaural signal reproduction, leaving a comprehensive design methodology unavailable.

In an effort to overcome the limitations of existing beamforming-based methods and to accurately reproduce binaural signals captured by arrays of arbitrary geometry, a number of techniques have been developed \cite{ref9,ref10,ref11,ref12,new1}. The binaural signal-matching (BSM) technique is one such technique \cite{ref13,ref46}.
BSM entails estimating the binaural signals directly from the array measurements using least-squares (LS) optimization. To enhance perception, at high frequencies, Magnitude Least-Squares (MagLS) optimization is used instead \cite{ref29}. A BSM design was recently presented for a semi-circular array with varying numbers of microphones, incorporating head tracking \cite{ref13}. The study revealed that BSM accuracy heavily depends on microphone positioning, leading to poor performance when a listener's ear is distant from the recording microphones. In addition, the BSM which is designed under a diffuse sound-field assumption may be sub-optimal under high Direct-to-Reverberant Ratio (DRR) conditions, for example \cite{ref32}.

Parametric spatial audio was also investigated as an alternative for binaural reproduction. In these approaches, the sound field is decomposed into direct and reverberant components, each estimated and reproduced independently \cite{ref22,ref24,ref25,ref26,ref27}. While this approach has shown promising results, its accuracy depends on the estimation of signal-dependent model parameters, such as the direction-of-arrival (DOA) of sound sources, and the direct-to-reverberant ratio (DRR) of the sound field. \textcolor{\MyColor}{This uncertainty in the performance of the parametric approach defines a current gap in BSM-based binaural reproduction, where the choice of what is preferable - the robustness of signal-independent BSM or the performance of signal-dependent BSM — may not be entirely clear.}  

This paper aims to address this gap by \textcolor{\MyColor}{systematically investigating the incorporation of signal information into BSM. To diversify the study, two signal-dependent methods are investigated.} The first incorporates signal information in the correlation matrix employed in the BSM design \cite{ref32}, and the second is based on parametric spatial coding, e.g. COMPASS \cite{ref27}, where the residual sound field complementing individual source modeling is reproduced using a BSM like approach. 

\textcolor{\MyColor}{Performance of these methods is investigated and compared to signal-independent BSM under a wide range of conditions that include a speaker in a reverberant room, with a range of source directions, direction estimations errors, and listener head rotations. The simulation study based on a semi-circular array mounted on a rigid sphere, is finally complemented by a listening test. Important conclusions arise from this study that can be useful when incorporating these methods:}
\textcolor{\MyColor}{
\begin{enumerate}
\item As expected, a clear improvement in performance is observed for the signal-dependent BSM methods, in particular under head rotations, where one ear is typically rotated away for the array. 
\item The improved performance at the source direction comes at only a marginal cost of a small reduction in performance across other directions.
\item When source direction estimation is poor, the signal-dependent approaches degrade in performance but only to match that of the signal-independent BSM.
\end{enumerate}
}

\textcolor{\MyColor}{
These results reinforce the confidence in the signal-dependent BSM approaches, showing relative robustness by not degrading below the signal-independent BSM under conditions of error in parameter estimation.
}

Note: the work is a significant extension of a conference paper by the same authors that investigated the feasibility of incorporating signal information into BSM \cite{ref47}.

\section{Mathematical Background}
\subsection{Signal Model}
\label{sec:Signal_Model}
This subsection provides the mathematical model of the signal as used in this paper. Throughout the paper, the standard spherical coordinate system is used, denoted by $(r,\theta,\phi)$, where $r$ is the distance to the origin, $\theta$ is the elevation angle measured from the Cartesian z axis downwards to the Cartesian xy plane, and $\phi$ is the azimuth angle measured from the positive x axis towards the positive y axis.
We denote $k=\frac{2\pi}{\lambda}$ as the wave-number, where $\lambda$ is the wave-length, $f$ is the frequency and $\Omega=(\theta,\phi)$ represents the arrival direction of a plane wave in the sound field. 
Further, assume that the sound-field is composed of $L$ far-field sound sources generating plane waves with arrival directions ${\Omega_l},{l=1,...,L}$, and source signals ${s_l(f)},{l=1,...,L}$ \cite{ref15}. The sound-field is captured by an array of $M$ microphones, which are located at ${(r_m,\Omega_m)},{m=1,...,M}$, centered at the origin. The noisy array measurements can therefore be described by the following narrow-band model \cite{ref15}:
\begin{equation}
    \mathbf{x}(f)=\mathbf{V}(f)\mathbf{s}(f)+\mathbf{n}(f)
    \label{eq:1}
\end{equation}
where ${\mathbf{x}(f)=[x_1(f),x_2(f),...,x_M(f)]^T}$ is the microphone-signal vector (measurements), $\mathbf{V}(f)=[\mathbf{v}(k,\Omega_1),\mathbf{v}(k,\Omega_2),...,\mathbf{v}(k,\Omega_L)]$ is an $M\times L$ complex matrix with columns $\mathbf{v}(k,\Omega_l)$ representing the array steering vector from the $l$-th source to the microphone positions for all ${l=1,...,L}$ sources, ${\mathbf{s}(f)=[s_1(f),s_2(f),...,s_L(f)]^T}$ is the source-signal vector, and $\mathbf{n}(f)$ is an additive-noise vector.

In the BSM model we further assume that a listener is virtually positioned with the center of the head coinciding with the origin,  where $h^{l,r}(k,\Omega)$ denotes the HRTF of the left and right ears of the listener using the superscripts $(\cdot)^l$ for the left ear and $(\cdot)^r$ for the right ear.
The signal at the left and right ears can now be written as \cite{ref40}:
\begin{equation}
\label{eq:binaural}
    p^{l,r}(f)=[\mathbf{h}^{l,r}(f)]^T\mathbf{s}(f)
\end{equation}
where $\mathbf{h}^{l,r}=[h^{l,r}(\Omega_1),h^{l,r}(\Omega_2),...,h^{l,r}(\Omega_L)]^T$ contains the HRTFs corresponding to the directions of the sources.\newline

\subsection{Binaural Signal Matching}
\label{sec:BSM}
This subsection provides the mathematical model of BSM. First, assume that the configuration of the microphone array is known, such that the steering matrix $\mathbf{V}(f)$ can be generated by simulation or measurement. In the first step, the array measurements are filtered and combined, in a similar manner to beamforming:
\begin{equation}
    \hat{p}^{l,r}(f)=[\mathbf{c}^{l,r}(f)]^H\mathbf{x}(f)
    \label{eqn:3}
\end{equation}
where $\mathbf{c}$ is an $M\times1$ complex vector holding the filter coefficients. Next, $\mathbf{c}$ is chosen to minimize the following mean-squared error between $\hat{p}^{l,r}(f)$ and $p^{l,r}(f)$, the binaural signals in (2), for each ear separately:
\begin{equation}
    err^{l,r}_{bin}(f)=\mathbb{E}[|p^{l,r}(f)-\hat{p}^{l,r}(f)|^2]
    \label{eq:2}
\end{equation}
where $\mathbb{E}[\cdot]$ is the expectation operator.
Next, assume that the noise is uncorrelated to the sources. This leads to the following error formulation \cite{ref13}:
\begin{equation}
\label{eqn:5}
    \begin{aligned}
    err^{l,r}_{bin}(f) &=\mathbb{E}[\|[\mathbf{s}(f)]^H([\mathbf{V}(f)]^H\mathbf{c}^{l,r}(f)-[\mathbf{h}^{l,r}(f)]^*)\|^2_2\\
    & +\|[\mathbf{n}(f)]^H\mathbf{c}^{l,r}(f)\|^2_2]
\end{aligned}
\end{equation}
Eq. (\ref{eqn:5}) can also be written in a matrix form, omitting the explicit dependence on frequency, as:
\begin{equation}
\label{eqn:5_matrix}
    \begin{aligned}
        err^{l,r} &=((\mathbf{c}^{l,r})^H\mathbf{V}-[\mathbf{h}^{l,r}]^T)\mathbf{R_s}(\mathbf{V}^H\mathbf{c}^{l,r}-[\mathbf{h}^{l,r}]^*) \\
        & +(\mathbf{c}^{l,r})^H\mathbf{R_n}\mathbf{c}^{l,r}
    \end{aligned}
\end{equation}
Minimizing the error in Eq. (\ref{eqn:5_matrix}) leads to the following solution \cite{ref46}:
\begin{equation}
        \mathbf{c}^{l,r}_{opt}=(\mathbf{V}\mathbf{R_s}\mathbf{V}^H+\mathbf{R_n})^{-1}\mathbf{V}\mathbf{R_s}[\mathbf{h}^{l,r}]^*
    \label{eqn:6}
\end{equation}
where $\mathbf{R_s}=E[\mathbf{s}\mathbf{s}^H]$, $\mathbf{R_n}=E[\mathbf{n}\mathbf{n}^H]$.
Next, if $\mathbf{R_s}$ is unknown, one can assume that the sound field is diffuse as in \cite{ref21}, leading to ${\mathbf{R_s}=\sigma_s^2\mathbf{I}_L}$ and with sources DOA uniformly spread around the sphere. Additionally, it is assumed that the noise is uncorrelated across microphones and identically distributed, such that ${\mathbf{R_n}=\sigma_n^2\mathbf{I}_M}$, where $\sigma_n^2$ is the noise variance, $\sigma_s^2$ is the source variance and $\mathbf{I}_M$,$\mathbf{I}_L$ are the $M\times M$ and $L\times L$ identity matrices, respectively. These assumptions lead to the following simplification of Eq. (\ref{eqn:6}) \cite{ref13}:
\begin{equation}
    \mathbf{c}^{l,r}_{BSM}=(\mathbf{V}\mathbf{V}^H+\frac{1}{SNR}\mathbf{I}_M)^{-1}\mathbf{V}[\mathbf{h}^{l,r}]^*
    \label{eqn:7}
\end{equation}
where $SNR=\frac{\sigma_s^2}{\sigma_n^2}$.

To reduce the negative impact of spatial cues at high frequencies becoming less accurate because of the loss of precision in BSM and increase in binaural error, it was suggested in \cite{ref21} to modify BSM at high frequencies by minimizing the magnitudes in Eq. (\ref{eqn:5}) using MagLS \cite{ref29}. It was based on the idea that ILD is more noticeable than ITD for spatial perception at high frequencies \cite{ref33}-\cite{ref34}.
In \cite{ref21} it was shown that even with the use of MagLS the BSM still has a high error for high frequencies and that the performance rapidly degrades the further the microphones are from the ears of a virtual listener. In addition, in cases where the sound-field is directional, such as a speaker in a room, the diffuse sound-field assumption made in the development of the BSM fails which may lead to even higher errors. In this paper, methods that aim to improve the BSM method, especially in cases of directional sound-fields will be presented and analysed. 

\section{BSM Approaches that Incorporate Signal Information}
\label{sec:GenSignalModel}
This section introduces two distinct BSM-based binaural reproduction approaches. These approaches incorporate signal information, specifically modeling dominant sound sources, extending beyond the solution provided in Eq. (\ref{eqn:7}), which assumes a diffuse field.
Throughout this section it is assumed that the sound-field, described in Eq. (\ref{eq:1}), is composed of direct and reverberant components. The direct component is composed of the direct sound from $D$ far-field sources corresponding to a DOA of $\mathbf{\Omega_d}=[{\Omega_1},...,{\Omega_D}]^T$. The reverberant component is approximated to be composed of $L$ far-field sources whose DOA's are uniformly spread on a sphere as in the model presented in Section \ref{sec:BSM}. This leads to the following formulation of the signal model:
\begin{equation}
\mathbf{x}=\mathbf{V_d}\mathbf{s_{d}}+\mathbf{V_r}\mathbf{s_{r}}+\mathbf{n}
\label{eq:new_model}
\end{equation}
where $\mathbf{V_d}$ is the $M\times D$ steering matrix of the direct component, $\mathbf{V_r}$ is the $M\times L$ assumed steering matrix for the reverberant component, and $\mathbf{s_{d}}$ and $\mathbf{s_{r}}$ are the source signal vectors for the direct and reverberant components, respectively.
It is further assumed that the measurement noise $\mathbf{n}$ is white, uncorrelated to the source signals and that $\mathbb{E}[\mathbf{n}\mathbf{n}^H]=\sigma_n^2\mathbf{I}_M$.
Throughout this section, the following parameters are assumed to be separately estimated:
\begin{itemize}
    \item The DOA of the $D$ direct sources, $\mathbf{\Omega_d}$. Throughout this section, the dependency on $\mathbf{\Omega_d}$ for the direct component is omitted for simplicity. 
    
    \item The measurements correlation matrix $\mathbf{R_x}=\mathbb{E}[\mathbf{x}\mathbf{x}^H]$.
\end{itemize}
It is also assumed that the source signals of the direct component are estimated by applying a linearly-constrained minimum-variance (LCMV) beamformer \cite{ref35}:
\begin{equation}
    \mathbf{\hat{s}_{d}}=\mathbf{W_d}\mathbf{x}
    \label{eq:sd}
\end{equation}
where $\mathbf{W_d}$ is the weights matrix calculated using the constraint $\mathbf{W_d}\mathbf{V_d}=\mathbf{I}_D$ where $\mathbf{I}_D$ is a $D\times D$ identity matrix, as shown in \cite{ref35} which leads to: 
\begin{equation}
\label{eq:23}
    \mathbf{W_d}=(\mathbf{V_d}^H(\mathbf{R_x})^{-1}\mathbf{V_d})^{-1}\mathbf{V_d}^H(\mathbf{R_x})^{-1}
\end{equation}
Note that because the LCMV is computed with correlation matrix $\mathbf{R_x}$, it is assumed that $\mathbf{s_{d}}$ is uncorrelated to $\mathbf{s_{r}}$ which is required to avoid signal cancellation \cite{ref41}. 
\subsection{BSM Incorporated with Parametric Spatial Coding}
\label{sec:BSM-Leo}
This subsection outlines a mathematical model that integrates a parametric spatial coding method with the BSM technique for an arbitrary array. This approach will be referred to as COMPASS-BSM (COM) \cite{ref48,ref49}.
In this approach it is assumed that the measured signal can be described by Eq. (\ref{eq:new_model}).
Using the estimated $\hat{\mathbf{s}}_{d}$, the binaural signal for the direct component can be calculated as follows Eq. (\ref{eq:binaural}):
\begin{equation}
    \hat{p}_d^{l,r}=[\mathbf{h}_d^{l,r}]^T\hat{\mathbf{s}}_{d}
    \label{eq:pd}
\end{equation}
where $\hat{p}_d^{l,r}$ represents the binaural signal for the direct component at the ears and $\mathbf{h}_{d}^{l,r}$ denote the HRTF vectors associated with the direct component.

With the estimated $\hat{\mathbf{s}}_{d}$, the reverberant plus noise component of the measurement can be estimated by substituting Eq. (\ref{eq:sd}) into Eq. (\ref{eq:new_model}), as:

\begin{equation}
    \mathbf{\hat{x}_r}=(\mathbf{I_M}-\mathbf{V_d}\mathbf{W_d})\mathbf{x}
    \label{eq:13}
\end{equation}

 Next, using $\mathbf{\hat{x}_r}$, the reverberant component of the binaural signal, $\hat{p}_r^{l,r}$, can be estimated using the BSM method as in Eq. (\ref{eqn:3}): 
\begin{equation}
    \hat{p}_r^{l,r}=(\mathbf{c}^{l,r}_{BSM})^H\mathbf{\hat{x}_r}
    \label{eq:pr}
\end{equation}
where $\mathbf{c}^{l,r}_{BSM}$ are the standard BSM weights which are calculate as shown in Section \ref{sec:BSM}.

Finally, the binaural signal for the entire sound field is calculated as the sum of the binaural signals for the direct and reverberant components:
\begin{equation}
    \hat{p}^{l,r}=\hat{p}^{l,r}_d+\hat{p}^{l,r}_r
    \label{eq:pleo}
\end{equation}
where the resulting filter weights can be written by substituting Eqs. (\ref{eq:pd}) and (\ref{eq:pr}) into (\ref{eq:pleo}), as:
\begin{equation}
    \mathbf{c}^{l,r}_{COM}(\mathbf{\Omega_d},f)=(\mathbf{I_M}-\mathbf{V_d}\mathbf{W_d})^H\mathbf{c}^{l,r}_{BSM}+\mathbf{W_d}^H[\mathbf{h}_{d}^{l,r}]^*
    \label{eq:c_COM}
\end{equation}
where $\mathbf{c}^{l,r}_{BSM}$ are defined in Eq. (\ref{eq:pr}), such that:
\begin{equation}
    \hat{p}^{l,r}=(\mathbf{c}^{l,r}_{COM})^H\mathbf{x}
\end{equation}

\subsection{BSM with Informed Correlation Matrix}
\label{sec:BSM-Shai}
In this subsection, a BSM-based binaural reproduction approach that incorporates the direct component information into the source correlation matrix is introduced with the goal of better estimating the optimal filter described in Eq. (\ref{eqn:6}). This approach will be referred to as Directional BSM (d-BSM).
In this approach it is assumed that the measured signal can be described by Eq. (\ref{eq:new_model}); that $\mathbf{s_{d}}$ and $\mathbf{s_{r}}$ are uncorrelated;  and that $\mathbb{E}[\mathbf{s_{r}}\mathbf{s_{r}}^H]=\sigma_r^2\mathbf{I}_L$, where $\mathbf{I}_L$ is the $L\times L$ identity matrix and $\sigma_r^2$ is the assumed variance of the sources belonging to the reverberant component. Note that the latter assumption holds when the reverberant component is diffuse.  
Using $\hat{\mathbf{s}}_{d}$ as in Eq. (\ref{eq:sd}) the covariance matrix of the sources belonging to the direct component can be estimated as:
\begin{equation}
    \mathbf{\hat{R}_{s_d}}=\mathbb{E}[\hat{\mathbf{s}}_{d}\hat{\mathbf{s}}_{d}^H]
    \label{eq:Rs}
\end{equation}
Using $\mathbf{\hat{x}_r}$ as in Eq. (\ref{eq:13}), the equivalent variance of the sources belonging to the reverberant component can be estimated as: 
\begin{equation}
    \hat{\sigma_r}^2=\frac{1}{L}\|\mathbf{\hat{x}_r}\|^2_2
    \label{eq:sigma_r}
\end{equation}
Next, the sources covariance matrix, $\mathbf{R_s}$ can be estimated from Eqs. (\ref{eq:Rs}) and (\ref{eq:sigma_r}) as:
\begin{equation}
\label{eq:Rs_new}
\mathbf{\hat{R}_s} = \begin{bmatrix}
\mathbf{\hat{R}_{s_d}} & \mathbf{0} \\
\mathbf{0} & \hat{\sigma_r}^2\mathbf{I}_L
\end{bmatrix}
\end{equation}
Now, by substituting $\mathbf{\hat{R}_s}$ in Eq. (\ref{eqn:6}), the optimal filter can be computed as:
\begin{equation}
\label{eq:10-new}
    \mathbf{c}^{l,r}_{d-BSM}(\mathbf{\Omega_d},f)=\mathbf{B}^{-1}(\mathbf{V_d}\mathbf{\hat{R}_{s_d}}[\mathbf{h}_{d}^{l,r}]^*+\hat{\sigma_r}^2\mathbf{V_r}[\mathbf{h}_{r}^{l,r}]^*)
\end{equation}
where $\mathbf{h}_{r}^{l,r}$,$\mathbf{h}_{d}^{l,r}$ denote the HRTF vectors associated with the reverberant and direct components, respectively, $\mathbf{c}^{l,r}_{d-BSM}(\mathbf{\Omega_d})$ is the filter weight vector and $\mathbf{B}$ is calculated as:
\begin{equation}
\mathbf{B}=\hat{\sigma_r}^2\mathbf{V_r}\mathbf{V_r}^H+\mathbf{V_d}\mathbf{\hat{R}_{s_d}}\mathbf{V_d}^H+\sigma^2_n\mathbf{I}_M
\end{equation}
As in the standard BSM, at high frequencies, MagLS \cite{ref29} can be used to minimise Eq. (\ref{eqn:5}). Observing Eq. (\ref{eqn:5}) it can be noted that it can equivalently be written as:
\begin{equation}
\label{eq:weight}
    \begin{aligned}
    err^{l,r}_{bin}(\mathbf{\Omega_d},f) &=\|\mathbf{{R}_s}^{0.5}([\mathbf{V}]^H\mathbf{c}^{l,r}-[\mathbf{h}^{l,r}]^*)\|^2_{2}\\
    & +\|\mathbf{{R}_n}^{0.5}\mathbf{c}^{l,r}\|^2_{2}
\end{aligned}
\end{equation}
Where $\mathbf{V}=[\mathbf{V_d},\mathbf{V_r}]$ and $\mathbf{h}^{l,r}=[\mathbf{h_d}^{l,r};\mathbf{h_r}^{l,r}]$.
By substituting Eq. (\ref{eq:Rs}) into Eq. (\ref{eq:weight}) the MagLS formulation can be written as:
\begin{equation}
\begin{aligned}
    \mathbf{c}^{l,r}_{d\text{-}BSM}(\mathbf{\Omega_d},f) &= \arg \min_{\mathbf{c}} \left\| |\mathbf{\hat{R}_s}^{0.5}\mathbf{V}^H\mathbf{c}^{l,r}| - |\mathbf{\hat{R}_s}^{0.5}[\mathbf{h}^{l,r}]^*| \right\|^2_2 \\
    &\quad + \sigma_n^2\|\mathbf{c}^{l,r}\|^2
\end{aligned}
\end{equation}
\subsection{Distinction between the two approaches}
\label{sec:dif}
This subsection aims to provide some insight into the differences between the two approaches.
The first method separates direct and reverberant components to provide a better model for the reverberant component but may be sensitive to estimation errors.
Conversely, the second method estimates only the variance of the reverberant component without needing to estimate this component's signal. It is therefore expected to be more robust against estimation errors.
Essentially, the trade-off between the two methods lies in the detail of modeling and robustness to estimation error.

\section{Error Measures}
\label{sec:err}
This section introduces the main error measures that are used throughout this paper to evaluate and compare the different binaural reproduction methods. 
\subsection{Normalised Mean-Square Error}
\label{sec:NMSE}
The normalized mean-squared error (NMSE) of the binaural signals is calculated as:
\begin{equation}
\label{eq:NMSE_LS}
    NMSE^{l,r}(f)=
    \begin{cases}
        \frac{|\hat{p}^{l,r}(f)-{p}^{l,r}_{ref}(f)|^2}{|{p}^{l,r}_{ref}(f)|^2}, & \text{for } f<1.5\text{kHz} \\
        \frac{||\hat{p}^{l,r}(f)|-|{p}^{l,r}_{ref}(f)||^2}{|{p}^{l,r}_{ref}(f)|^2}, & \text{for } f\geq1.5\text{kHz}
    \end{cases}
\end{equation}

where ${p}^{l,r}_{ref}(f)$ is the reference binaural signal of a certain frequency, and $\hat{p}^{l,r}(f)$ is the reproduced binaural signal of the same frequency.
For frequencies above $1.5\,$kHz the NMSE of the absolute value was calculated based on magnitude only, aligning with sound perception, as shown in \cite{ref21}.

\subsection{ITD and ILD}
\label{sec:ITD&ILD}

In this subsection, performance measures based on ITD and ILD are outlined. Throughout this section, it is assumed that the sound-field, as described in Section \ref{sec:Signal_Model}, comprises solely of a single source, having a DOA denoted as $\Omega$, with a unit variance. Using Eq. (\ref{eqn:3}), both $p^l(t)$ and $p^r(t)$, representing the left and right time-domain binaural signals, were computed for a chosen BSM filter. The reference binaural signals were derived by substituting the source signal in Eq. (\ref{eq:binaural}).

The ITD computation employed a well-accepted perceptual measure \cite{ref36}, utilizing a cross-correlation based method. Initially, the binaural signals were low-pass filtered at a cut-off frequency of $1.5\,$kHz, followed by the ITD calculation depicted as:
\begin{equation}
    ITD(\Omega)=\argmax_\tau|\sum_{t=0}^{T-\tau-1}p^r(t)p^l(t+\tau)|
\end{equation}
where $T$ represents the total number of time samples. The chosen cutoff frequency of $1.5\,$kHz aligns with the frequency range crucial for spatial perception through phase or binaural time differences.

After computing ITD for both the BSM filter and the reference signal, the ITD error measure was computed as:
\begin{equation}
    \epsilon_{ITD}(\Omega)=|ITD(\Omega)-ITD_{ref}(\Omega)|
\end{equation}

For the ILD estimation, the binaural signals was processed using ERB filter bands, as outlined in \cite{ref37}:
\begin{equation}
    ILD_f(f_c,\Omega)=10\log_{10}\frac{\sum_{f=0}^{f_c^{max}}|G(f,f_c)p^l(f)|^2}{\sum_{f=0}^{f_c^{max}}|G(f,f_c)p^r(f)|^2}
\end{equation}
where $G(f,f_c)$ represents the ERB filter with central frequency $f_c$ at frequency $f$, and $f_c^{max}$ denotes the maximum frequency of this ERB filter. An average of 22 filter bands in the $[1.5,20]\,$kHz range, generated using the Auditory Toolbox \cite{ref38}, was used to compute $ILD$ as:
\begin{equation}
    ILD(\Omega)=\frac{1}{22}\sum_{f_c}ILD_f(f_c,\Omega)
\end{equation}
Subsequently, the ILD error measure was computed as:
\begin{equation}
    \epsilon_{ILD}(\Omega)=\frac{1}{22}\sum_{f_c}|ILD_f(f_c,\Omega)-ILD_{ref}(f_c,\Omega)|
\end{equation}

\section{Simulation Study}
In this section, a comparison between the two approaches, presented in Section \ref{sec:GenSignalModel}, and the standard BSM, presented in Section \ref{sec:BSM}, is provided. Broadly, this section aims to study the benefits and to gain insights into signal-dependent BSM approaches 
\textcolor{\MyColor}{under various conditions, including robustness to estimation error.}

\subsection{Simulation Set-Up}
\label{sec:set-up}
In this section, a semi-circular microphone array, mounted on a rigid sphere, comprising $M=6$ omnidirectional microphones arranged on the horizontal plane was simulated. The semi-circular array was chosen since it is a simplified representation of a head-worn array, such as augmented reality glasses \cite{new2}. The microphone positions were denoted using spherical coordinates $(r,\frac{\pi}{2},\phi_m)$ for $m=1,\ldots,M$ relative to the array center, where $r=10\,$cm and $\phi_m=\frac{\pi}{2}-\frac{\pi(m-1)}{M-1}\,$rad. The steering vectors of this array were computed analytically in the SH domain following section 4.2 in \cite{ref17} using the array geometry and the DOAs of the far-field sources modeled as plane waves.
The HRTF in the simulation was obtained from the Cologne database measurements of the Neumann KU100 manikin \cite{ref18}. The HRTFs for the DOAs of the assumed sources were interpolated in the SH domain using a SH order of 30. 
The assumed listener head was aligned with the positive $x$ axis. An illustration of the array position relative to the head position and alignment is presented in Figure \ref{fig:illustration1}. An orientation of $90^\circ$ as shown in Figure \ref{fig:subfig2illus} of the array relative to the head was found to be the most challenging for the BSM algorithm \cite{ref21} since the right ear is relatively far from any of the arrays microphones compared to the standard case of no head rotation as shown in \ref{fig:subfig1illus}.

\begin{figure}[htbp]
    \centering
    \subfigure[]{\includegraphics[width=2.5cm]{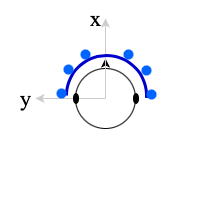}\label{fig:subfig1illus}}
    \subfigure[]{\includegraphics[width=2.5cm]{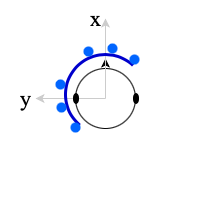}\label{fig:subfig3illus}}
    \subfigure[]{\includegraphics[width=2.5cm]{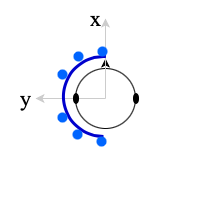}\label{fig:subfig2illus}}
    \caption{An illustration of a virtual listener head and the semi-circular array assuming a: \ref{fig:subfig1illus} $0^\circ$ array rotation. \ref{fig:subfig3illus} $50^\circ$ array rotation.  \ref{fig:subfig2illus} $90^\circ$ array rotation. The blue dots represent the array microphones, and the $x$ and $y$ grey arrows represent the positive $x$ and $y$ axes, respectively.}
    \label{fig:illustration1}
\end{figure}

Next, three different scenarios were simulated. In all three scenarios a point source was simulated inside a room using the image method \cite{ref19}. The semi-circular microphone array was centered at a position inside the room facing the $x$ axis as in Figure \ref{fig:subfig1illus}. The remaining details of the three simulated scenarios are presented In Table \ref{en:sims}. 

\begin{table}[ht]
\centering
\begin{tabular}{|c|c|c|c|}
\hline
 & \textbf{Scenario 1} & \textbf{Scenario 2} & \textbf{Scenario 3} \\
\hline
\textbf{Source} & Female & Male & Castanets \\
\textbf{Signal} & Speech & Speech &  \\
\hline
\textbf{Duration (s)} & 5 & 10 & 7.5 \\
\hline
\textbf{Array Position (m)} & $(4,3,1.7)$ & $(1.5,2,1.7)$ & $(4,3,1.7)$ \\
\hline
\textbf{Source Distance from} & $0.6$ & $0.9$ & $0.8$ \\
\textbf{Array (m)} & & & \\
\hline
\textbf{Room Dimensions (m)} & $6\times 4\times 3$ & $7\times 5\times 3$ & $8\times 6\times 4$ \\
\hline
\textbf{$T_{60}$ (s)} & $0.69$ & $0.65$ & $0.64$ \\
\hline
\end{tabular}
\caption{Simulation Parameters}
\label{en:sims}
\end{table}

 Scenario's 1 and 2 source signals were taken from the TIMIT database \cite{ref20}, and all signals were sampled at $48\,$kHz.

\subsection{Methodology}
\label{sec:methodology}
In this study, $\mathbf{c}^{l,r}_{BSM}$ was computed following Eq.(\ref{eqn:7}) as detailed in Section \ref{sec:BSM}, and MagLS \cite{ref29} was employed for frequencies within the range of $[1.5,24]\,$kHz. The steering matrix was computed assuming $L=400$ sources nearly-uniformly distributed on a sphere according to \cite{ref16}, as elaborated in Section \ref{sec:BSM}. The MagLS solution was obtained using the variable exchange method \cite{ref39} with an initial phase of $90^\circ$, a tolerance of $10^{-20}$, and a maximum of $10^5$ iterations. The steering vector of the direct component with a DOA of ${\Omega_d}$ was computed from the source and array positions (see Table \ref{en:sims}). 
Subsequently, after generating microphone signals, $\mathbf{x}(t)$, as described in section \ref{sec:set-up}, $\mathbf{x}[t,f]$ was computed using Short-Time Fourier Transform (STFT) with a Bartlett window of duration $0.032\,$s and a hop size of $0.008\,$s. Following this, the measurements' correlation matrix $\mathbf{R_x}$ was computed by averaging across both time and frequency domains according to:

\begin{figure}[t]
\centering
\subfigure[Right ear]{\includegraphics[width=9.5cm]{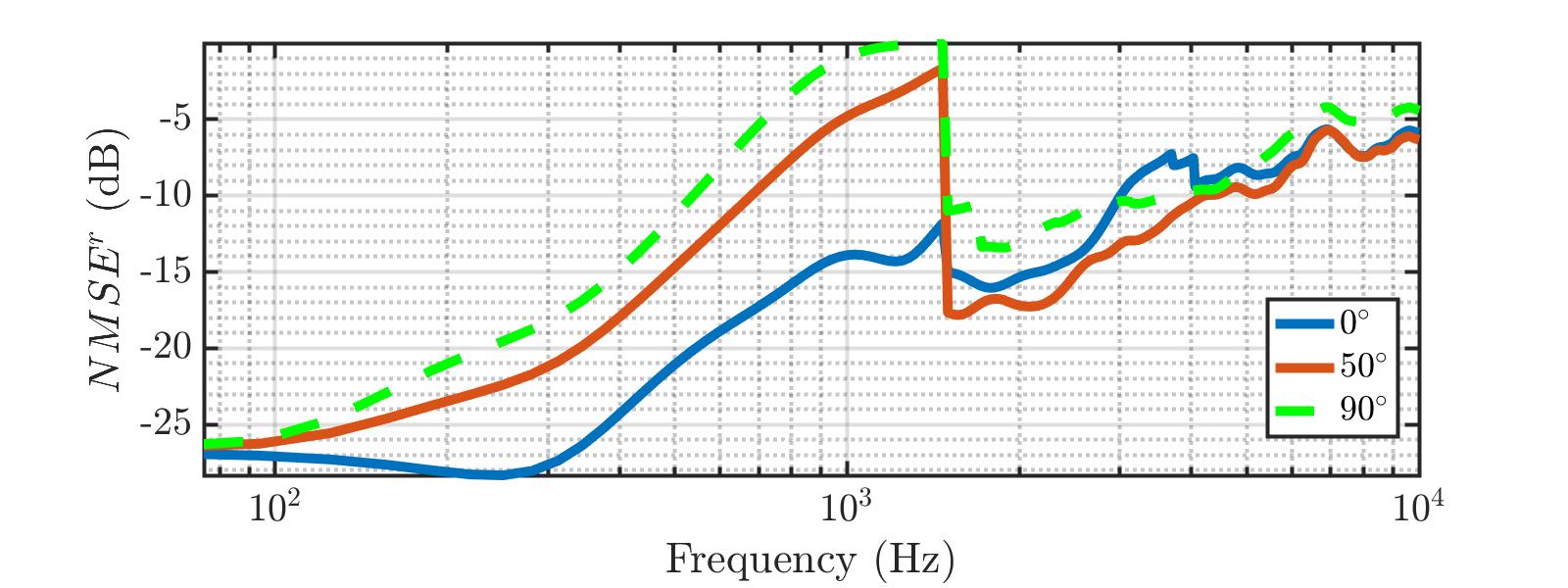}\label{fig:NMSE_R}}
\subfigure[Left ear]{\includegraphics[width=9.5cm]{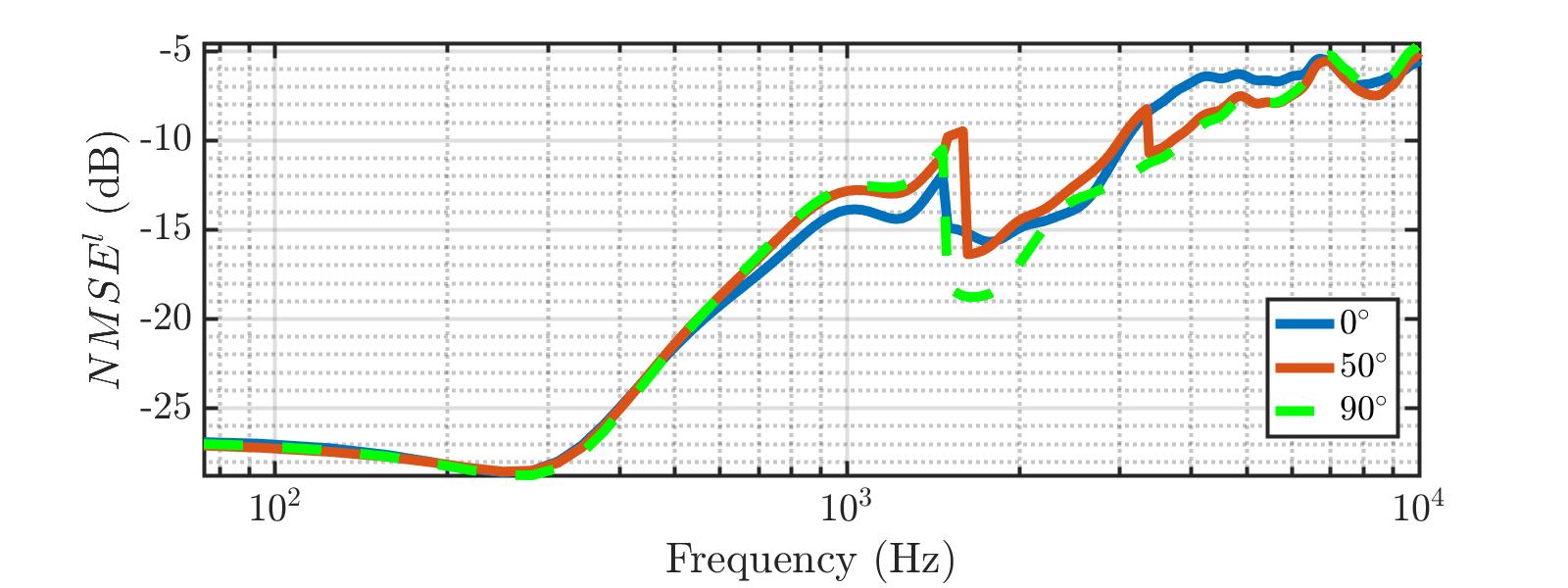}\label{fig:NMSE_L}}
\caption{The NMSE of the BSM method is presented in the top and bottom figures for the right and left ears, respectively. The NMSE values are computed as defined in Section \ref{sec:NMSE}, with reference to a diffuse sound field as detailed in \cite{ref21}. The evaluations consider an $SNR$ of $20\,$dB and three head-rotation of $90^\circ$, $50^\circ$ and no head rotation.}
\label{fig:NMSE}
\end{figure} 

\begin{equation}
    \mathbf{\hat{R}_x}[f]=\frac{1}{T}\sum_{t=1}^{T}\sum_{j=-J}^{J}\mathbf{x}[t,f+j\cdot \Delta f]\mathbf{x}^H[t,f+j\cdot \Delta f]
\end{equation}
where $T$ denotes the total length in seconds of the recording, $\Delta f\,$ is the frequency resolution and $J$ controls the bandwidth of the frequency smoothing.
Subsequently, the LCMV beamformer was computed following Eq. (\ref{eq:23}), leading to the determination of $\mathbf{c}^{l,r}_{COM}(\mathbf{\Omega}_d,f)$ using $\mathbf{c}^{l,r}_{BSM}$, as per Eq. (\ref{eq:c_COM}) in Section \ref{sec:BSM-Leo}. Further, $\mathbf{c}^{l,r}_{d-BSM}(\mathbf{\Omega}_d,f)$ was calculated following the equations outlined in Section \ref{sec:BSM-Shai}, employing the BSM steering matrix as the reverberant component's steering matrix and utilizing MagLS \cite{ref29} for frequencies within the range of $[1.5,24],$kHz.

Following the computation of the three filters, the binaural signals for all filters were computed using Eq. (\ref{eqn:3}) by replacing the appropriate filter $\mathbf{c}^{l,r}$. 
The reference binaural signals were obtained by convolving the HRTFs of the left and right ears with the HOA signals of order 14, which were computed using the image method as described in Section \ref{sec:set-up}.
Finally, the errors measures as presented in Section \ref{sec:err} were computed using the binaural signals for the three methods, and the reference binaural signals.

\subsection{Analysis of ITD and ILD}
\label{sec:analysisITD&ILD}
\begin{figure}[ht]
\centering
\subfigure[]{\includegraphics[width=9cm]{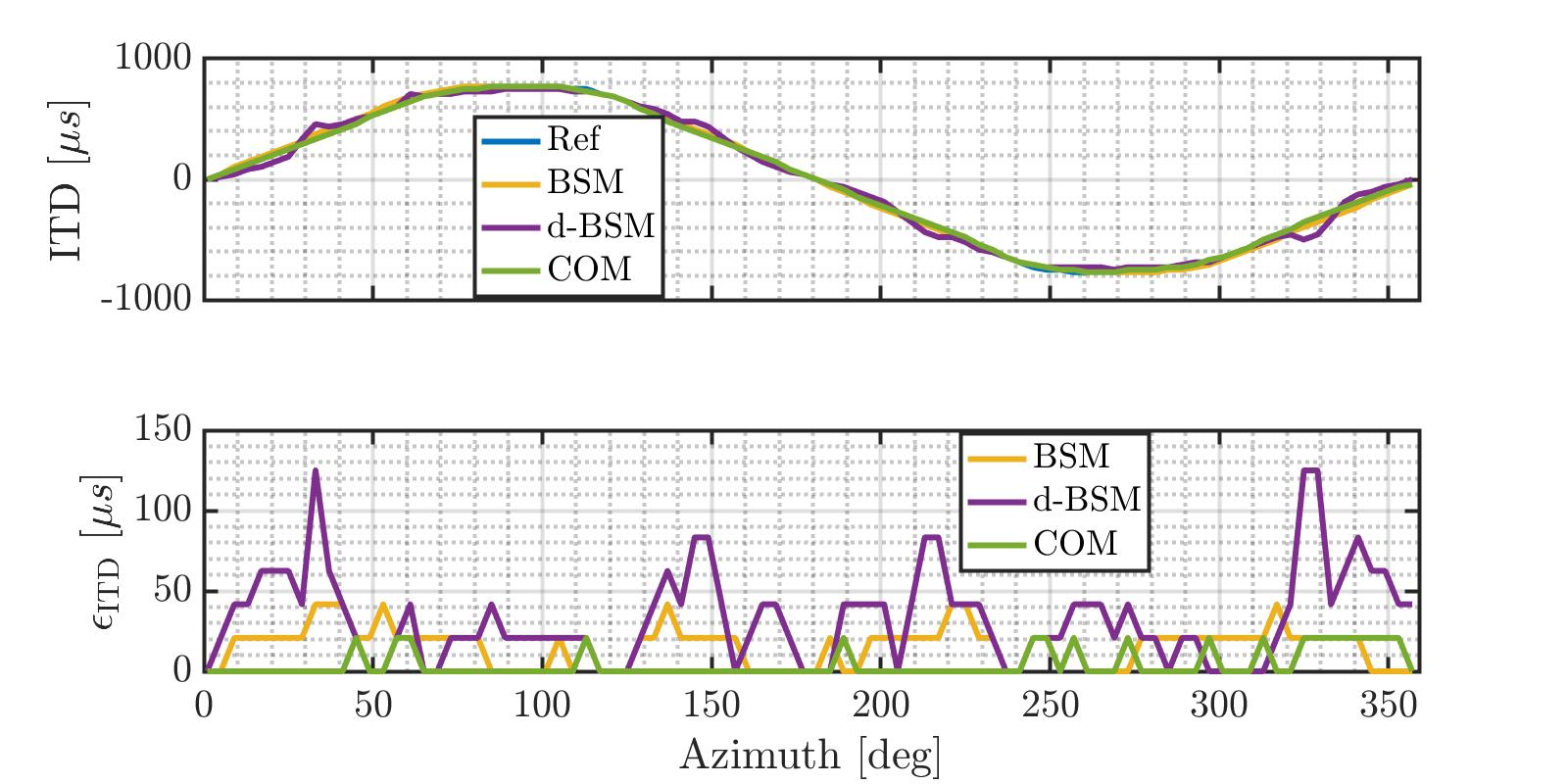}\label{fig:ITD_0}}
\subfigure[]{\includegraphics[width=9cm]{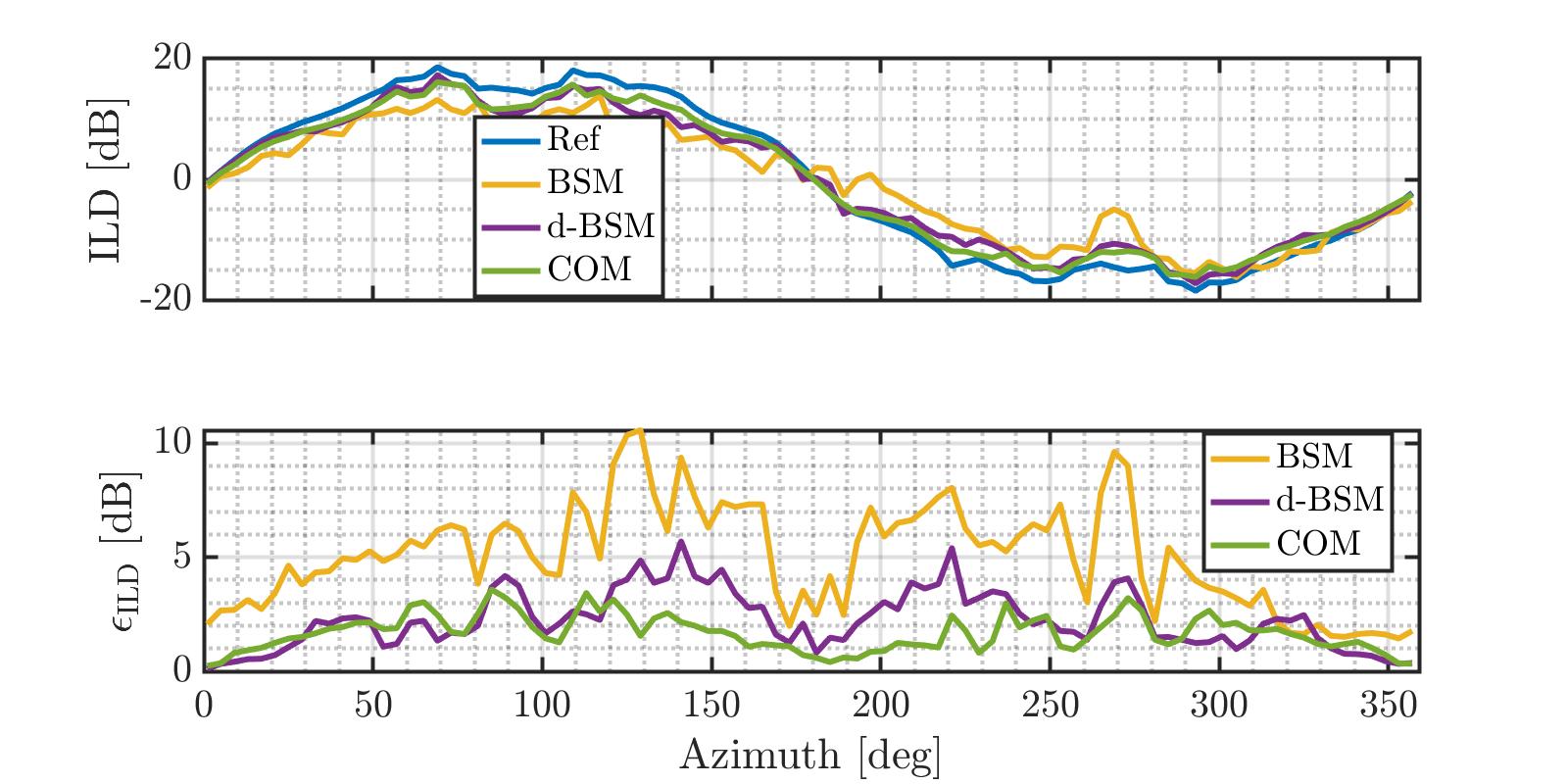}\label{fig:ILD_0}}
\caption{The ITD (a) and ILD (b) evaluated using the direct sound component of scenario 1 in Table \ref{en:sims}. The orange line represents the BSM method, the purple line corresponds to the d-BSM approach, and the green line depicts the COM approach. ITD and ILD errors are also presented, as defined in Section \ref{sec:ITD&ILD}. No head rotation was employed.}
\label{fig:sweep_0}
\end{figure}
Performance metrics studied thus far are rooted in the Mean Squared Error (MSE). However, this might not fully capture human perception. Therefore, a study employing perceptually driven measures, ITD and ILD, is introduced.
The ILD and ITD, as outlined in Section \ref{sec:ITD&ILD}, are presented in this section, for scenarios detailed in Table \ref{en:sims} comparing the different methods. 

The calculation of the ILD and ITD can be divided into two parts. First, a source in a room was simulated generating a scenario in Table \ref{en:sims}. The recorded signals at the array were then used to compute filters for the methods as described in Section \ref{sec:methodology}. Now, using the same filters, binaural signals were computed but this time for a sound field composed of a single source in free-field, with the same direction of the source in the room in Table \ref{en:sims}, and ILD and ITD were calculated as defined in Section \ref{sec:ITD&ILD}, for this specific source direction. Finally, this same process of computing filters and then ILD and ITD, was repeated for a range of source directions to complete the ILD and ITD analysis. This somewhat non-standard analysis was employed so that the filters could be computed using realistic signals, but then evaluated in terms ILD and ITD in the direction of the source only, as the direct sound is known to dominate sound localization.

Figure \ref{fig:sweep_0} illustrates the performance of ILD and ITD as detailed above for the three distinct methods in scenario 1, as outlined in Table \ref{en:sims}. The source direction was set on the horizontal plane with an elevation of $90^\circ$, and no rotation of the head was considered.
The figure shows that all three methods perform reasonably well in terms of ITD, and for most directions remain below the just notable difference (JND) of $100\mu s$ \cite{ref43}. 
Furthermore, both novel approaches outperformed the BSM method in terms of ILD with the COM method performing the best. However, for all methods error is mostly above the JND of 1dB \cite{ref43}. 

Figure \ref{fig:ILD_ITD_dir_female} illustrates the performance of ILD and ITD under the same conditions of Figure \ref{fig:sweep_0}, but with $50^\circ$ head rotation. The figure shows that ILD and ITD error increased significantly for the BSM method and insignificantly for the other two methods. 
This shows the superior performance of the d-BSM and COM methods in terms of ITD and ILD when considering head rotation. This can be explained by the explicit information of the source direction embedded in these two methods.

\begin{figure*}[!ht]
\centering
\subfigure[Scenario 1]{
    \includegraphics[width=9cm]{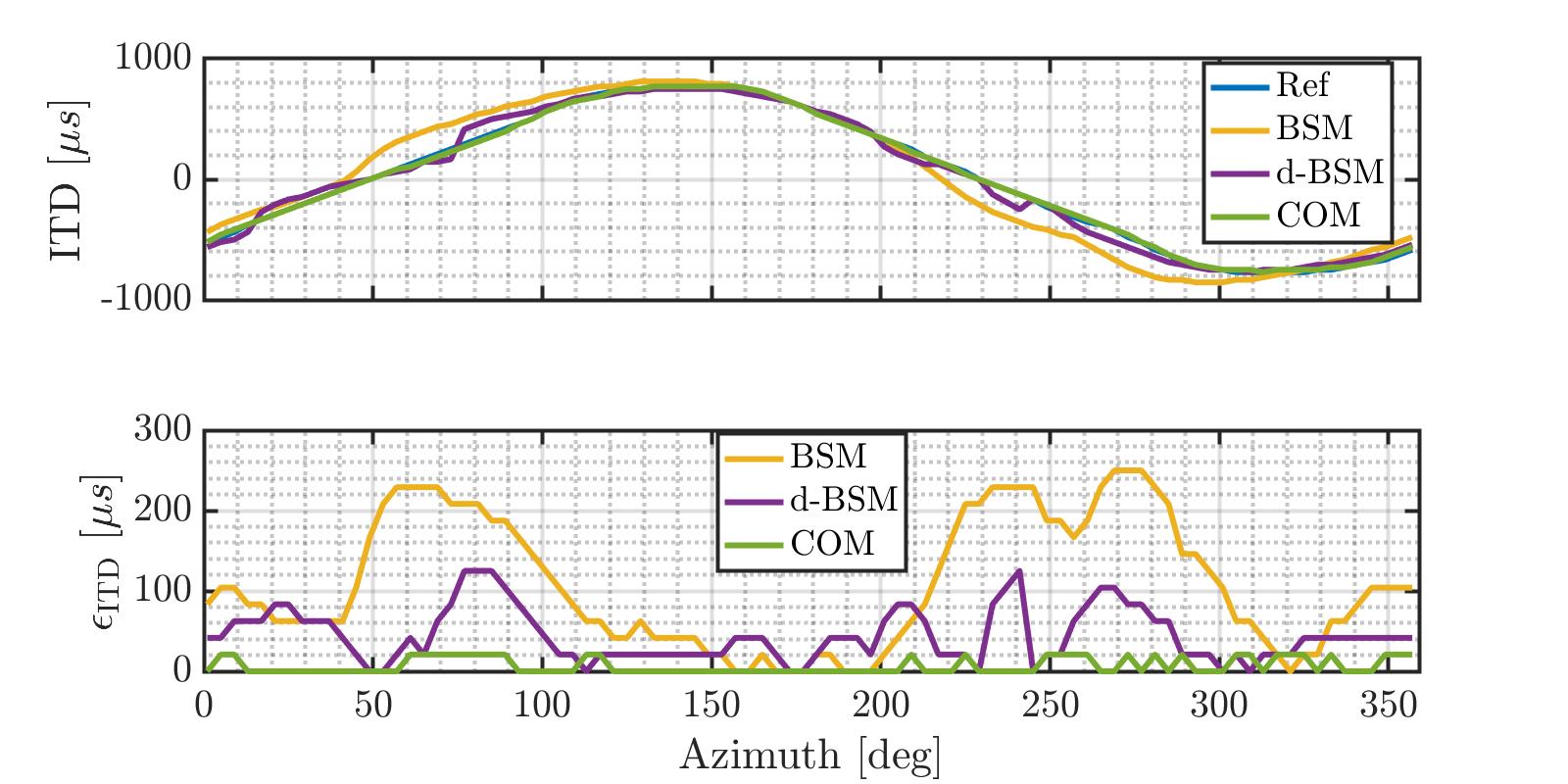}
    \includegraphics[width=9cm]{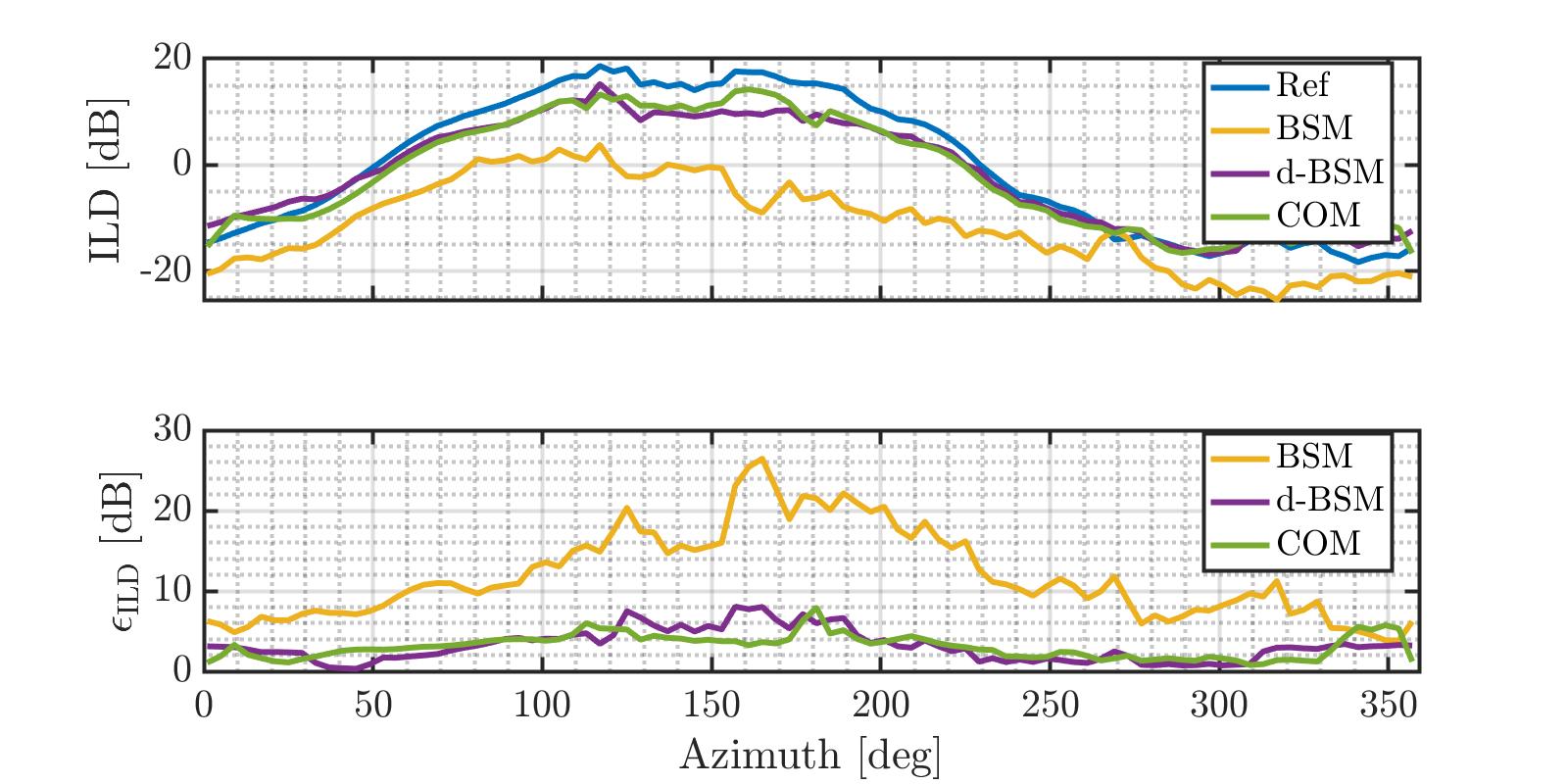}
    \label{fig:ILD_ITD_dir_female}
}
\subfigure[Scenario 2]{
    \includegraphics[width=9cm]{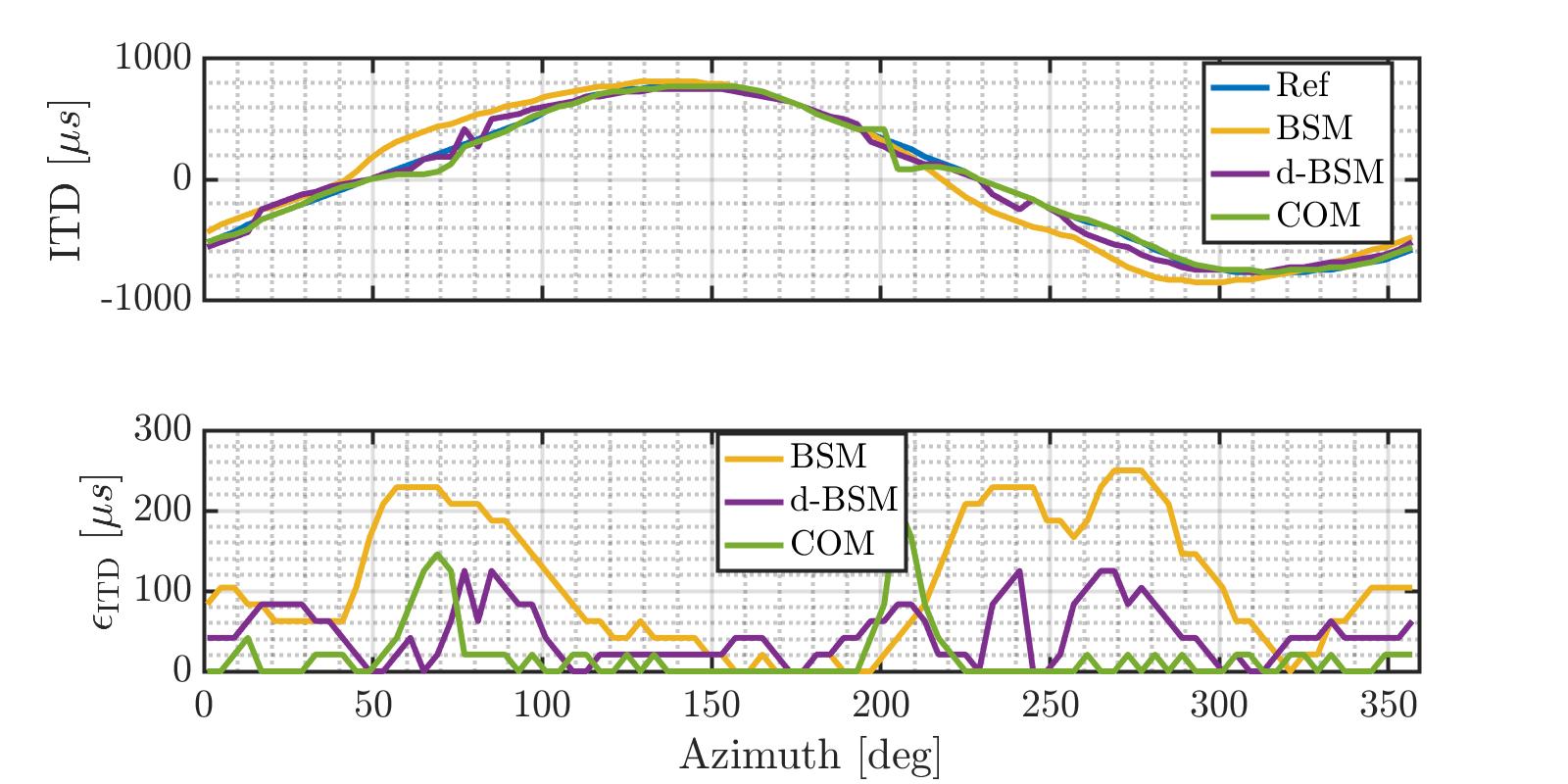}
    \includegraphics[width=9cm]{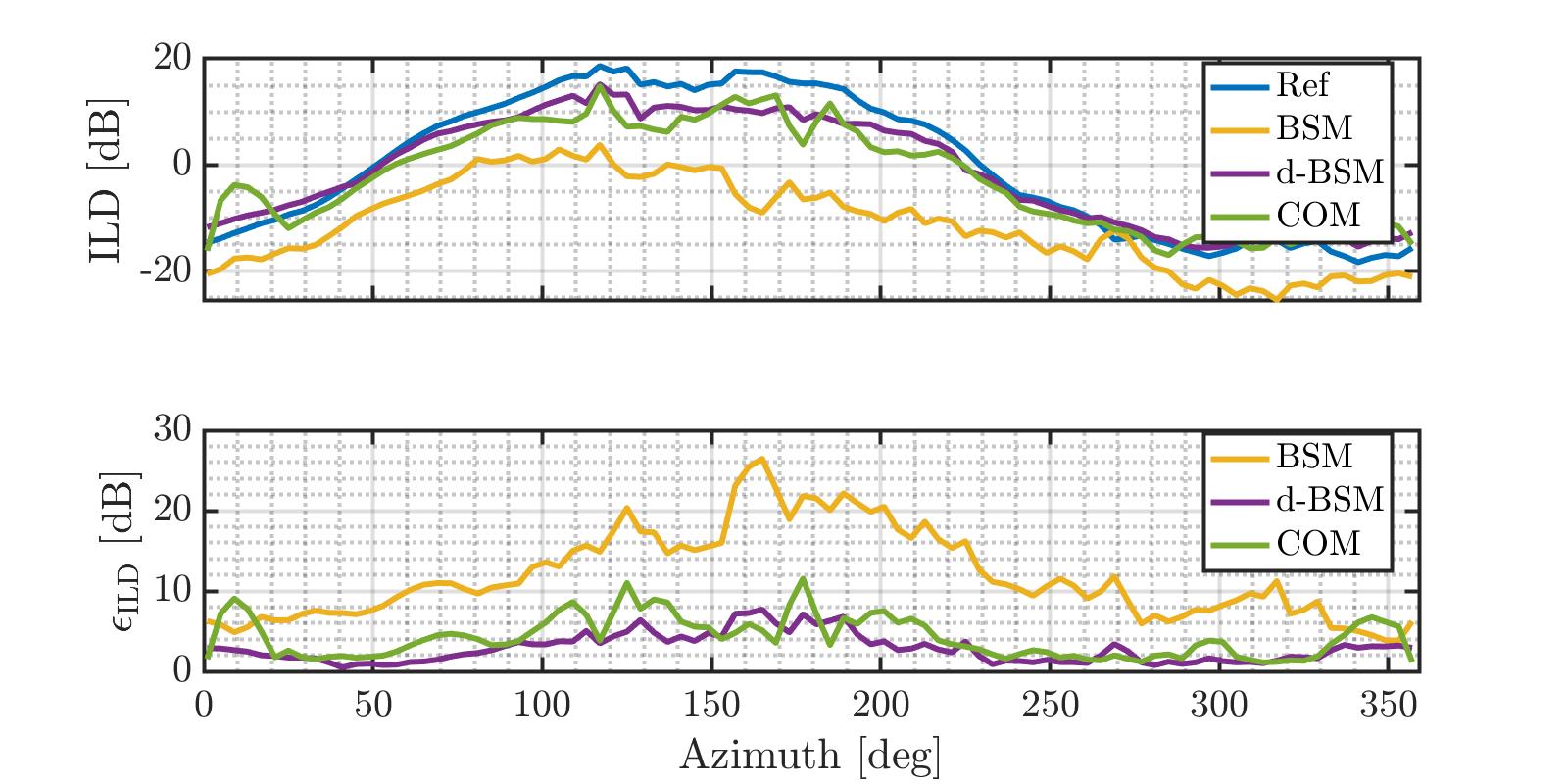}
    \label{fig:ILD_ITD_dir_male}
}
\subfigure[Scenario 3]{
    \includegraphics[width=9cm]{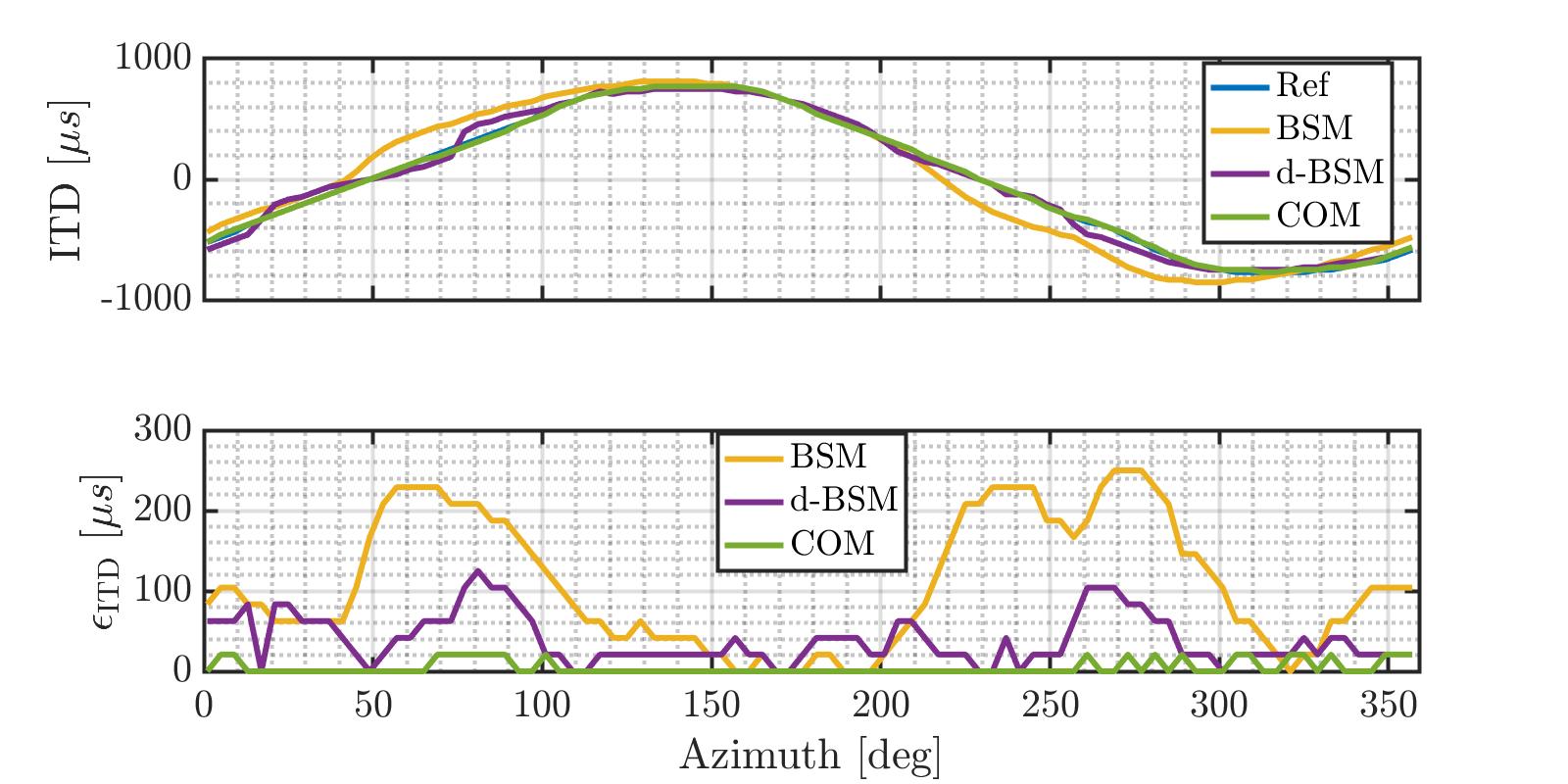}
    \includegraphics[width=9cm]{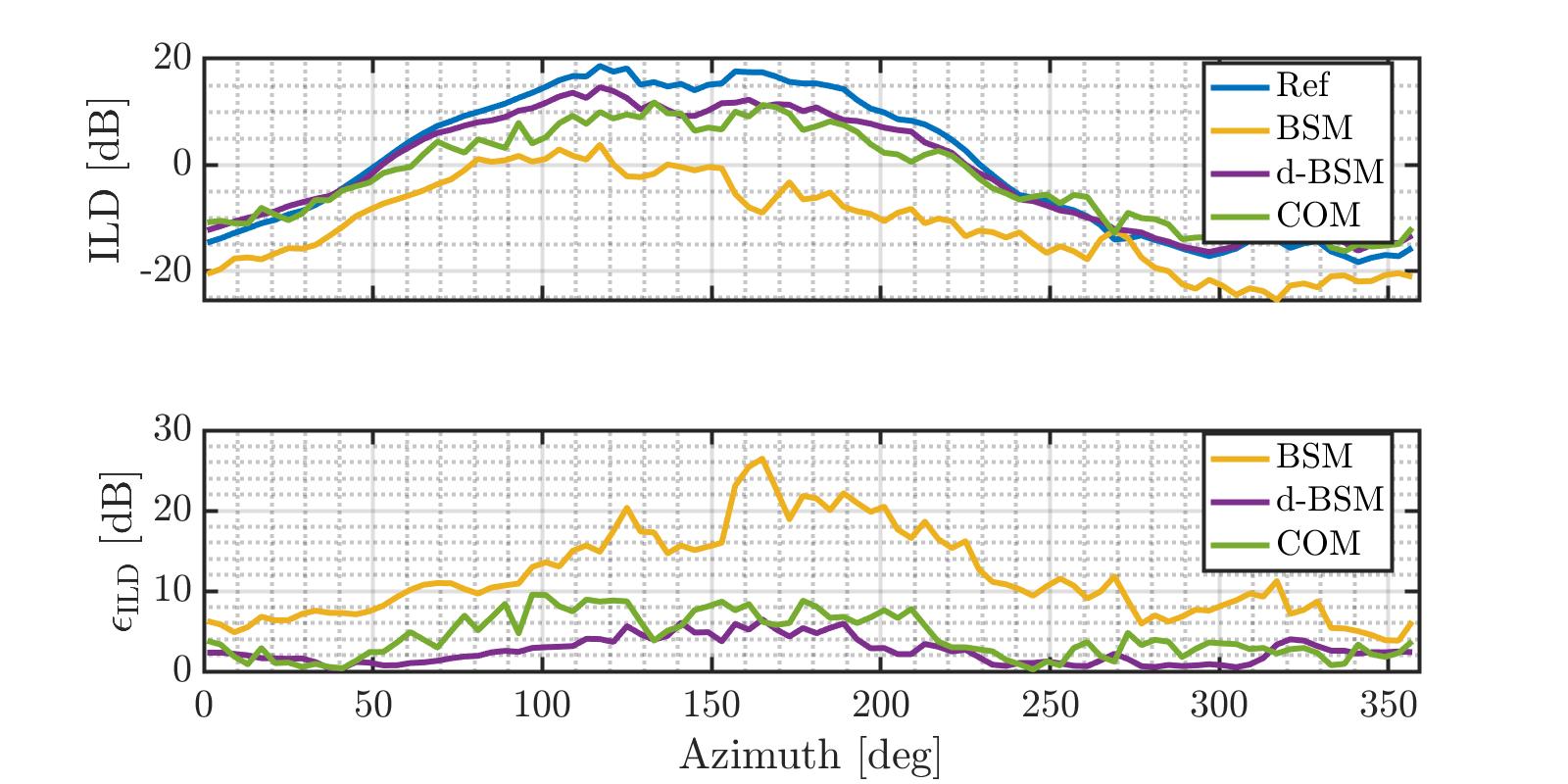}
    \label{fig:ILD_ITD_casta}
}
\caption{The ITD on the left and ILD on the right evaluated using the direct sound component of the three different scenarios described in Table \ref{en:sims}. The orange line represents the BSM method, the purple line corresponds to the d-BSM approach, and the green line depicts the COM approach. ITD and ILD errors are also presented, as defined in Section \ref{sec:ITD&ILD}. A $50^\circ$ head rotation was employed.}

\label{fig:ILD_ITD_50}
\end{figure*}

Figures \ref{fig:ILD_ITD_50} illustrates the performance of ILD and ITD under the same conditions of Figure \ref{fig:ILD_ITD_dir_female}, but for the different scenarios outlined in Table \ref{en:sims}. As shown in Figures \ref{fig:ILD_ITD_50}, the performance of the different approaches remains consistent across the various scenarios. This result helps to generalize the results of the adaptive filters shown in this section.

\subsection{Robustness to DOA Estimation Errors}
\label{sec:Robus}
In this subsection the ITD and ILD performance will be evaluated assuming an error in the estimation of source direction. Maintaining performance under such conditions is important as some estimation error is expected in practice. Assuming ${\Omega_d}$ is the true DOA of the source, the methods under study will be employed assuming the source has a DOA of ${\Omega_d}+{\Omega_{err}}$ where ${\Omega_{err}}$ is the assumed error in the DOA estimation.

\begin{figure}[ht]
\centering
\subfigure[]{\includegraphics[width=9cm]{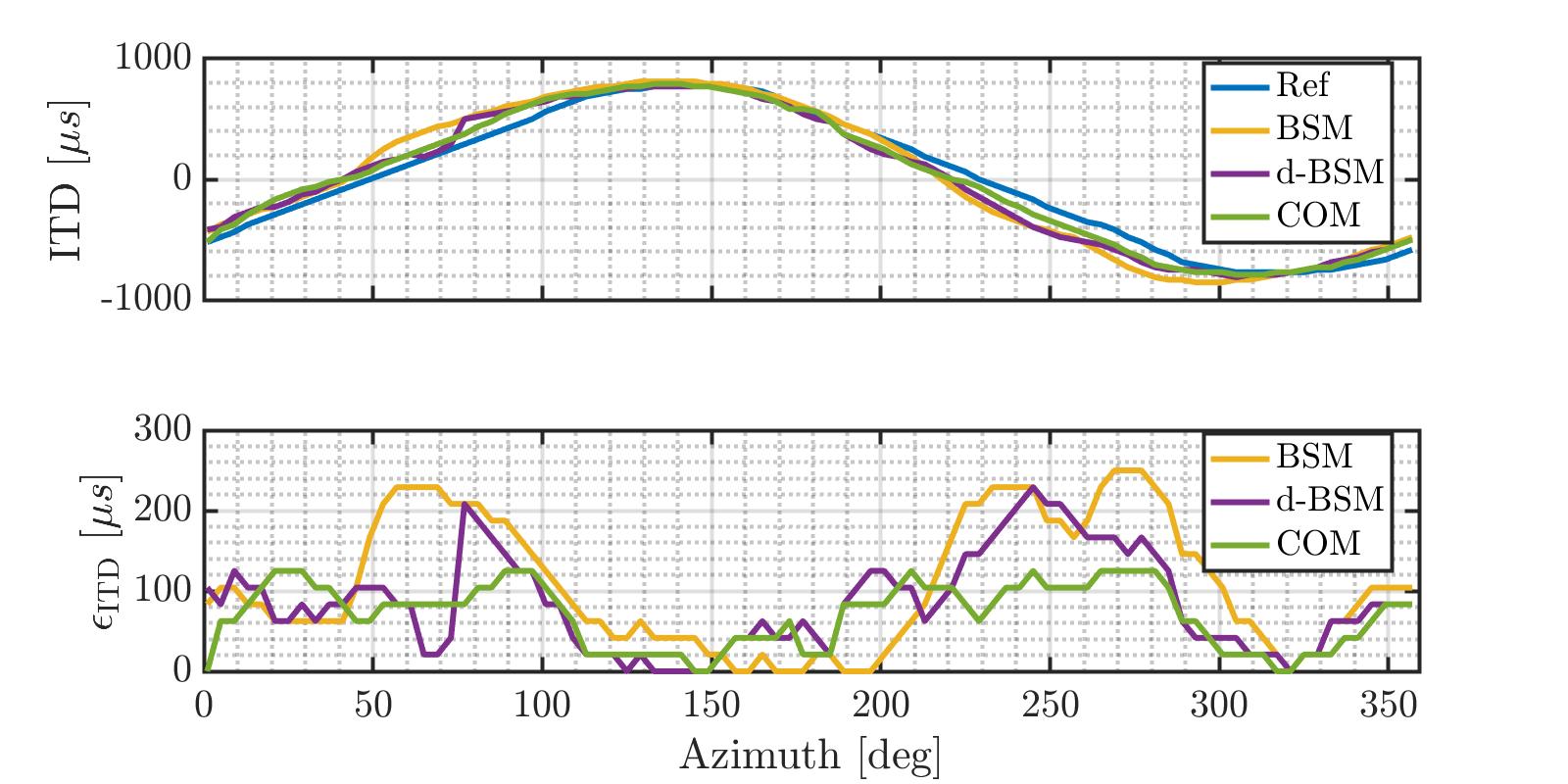}\label{fig:ITD_sweep_err_10ph5th_50}}
\subfigure[]{\includegraphics[width=9cm]{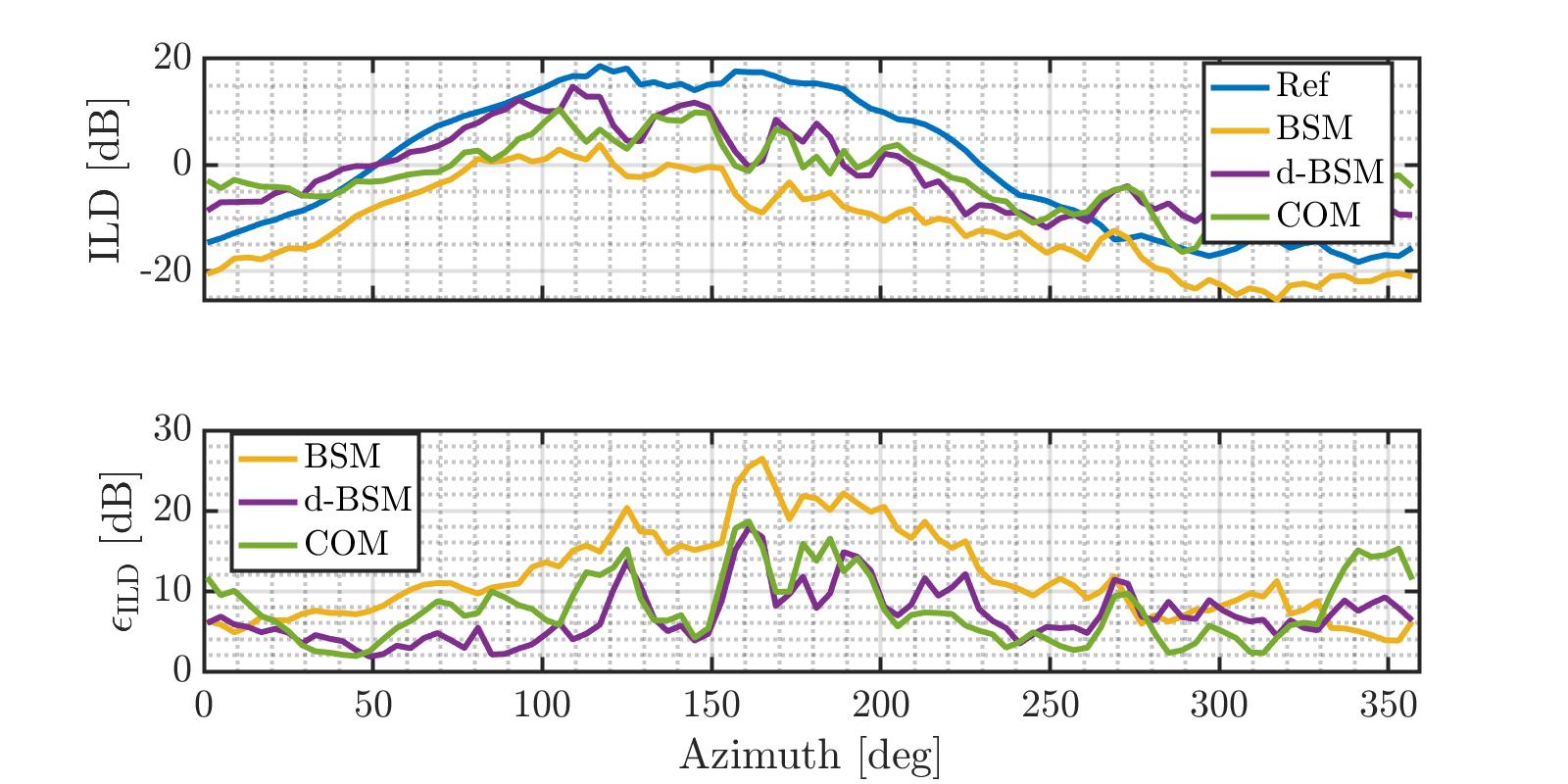}\label{fig:ILD_sweep_err_10ph5th_50}}
\caption{The ITD (a) and ILD (b) evaluated using the direct sound component of scenario 1 in Table \ref{en:sims}. The orange line represents the BSM method, the purple line corresponds to the d-BSM approach, and the green line depicts the COM approach. ITD and ILD errors are also presented, as defined in Section \ref{sec:ITD&ILD}. An error of ${\Omega_{err}}=(\phi=10^\circ,\theta=0^\circ)$ and a $50^\circ$ head rotation were employed.}

\label{fig:sweep_50_10ph5th_err}
\end{figure}

Figures \ref{fig:sweep_50_10ph5th_err} presents the ILD and ITD for the methods under study, assuming an elevation of $90^\circ$ with $50^\circ$ head rotation. All other conditions are the same as in Figure \ref{fig:ILD_ITD_dir_female}. An error of ${\Omega_{err}}=(\phi=10^\circ,\theta=0^\circ)$ was assumed, i.e. $10^\circ$ in azimuth.
By comparing Figures \ref{fig:sweep_50_10ph5th_err} to Figures \ref{fig:ILD_ITD_dir_female}, a clear degradation in ITD and ILD performance for both d-BSM and COM is observed. It can be observed that the d-BSM method is degraded less in its ITD performance compared to COM, and now both methods perform similarly. In addition it can be observed that for both d-BSM and COM methods, the ITD and ILD performance is better than the performance of the BSM method. However, the margin in performance between these two methods and the BSM has been reduced due the assumed DOA error. This can be explained by the robustness of the BSM method that doesn't rely on any DOA estimation. The degradation in performance of the d-BSM and COM can be explained by observing Eq. (\ref{eq:sd}). As this equation represents beamforming, and assuming the look direction is incorrect, the beamformer will produce a degraded estimation of the source signal. 

\subsection{Performance at directions away from source DOA}
In this subsection, the ILD and ITD performance of the different methods will be calculated for a specific case. In this case, a single source DOA is used for computing the filters, but analysis is performed in all azimuth directions, i.e. also away from source DOA. This analysis is important since it compares the performance of the different methods in directions other than source DOA, which may also be important, e.g. due to an onset of a source in these directions.
The calculation of the ILD and ITD in this chapter can also be divided into two parts.
First, a source in a room was simulated generating scenario 1 in Table \ref{en:sims}. The recorded signals at the array were then used to compute filters for the methods as described in Section \ref{sec:methodology}. In this case, the source was placed at ${\Omega}=(40^\circ,90^\circ)$ relative to the array. Now, using the same filters, binaural signals were computed and ILD and ITD were calculated as defined in Section \ref{sec:ITD&ILD}, for sound sources in free-field scanning all azimuth directions, and an elevation of $90^\circ$. 

\begin{figure}[ht]
\centering
\subfigure[]{\includegraphics[width=9cm]{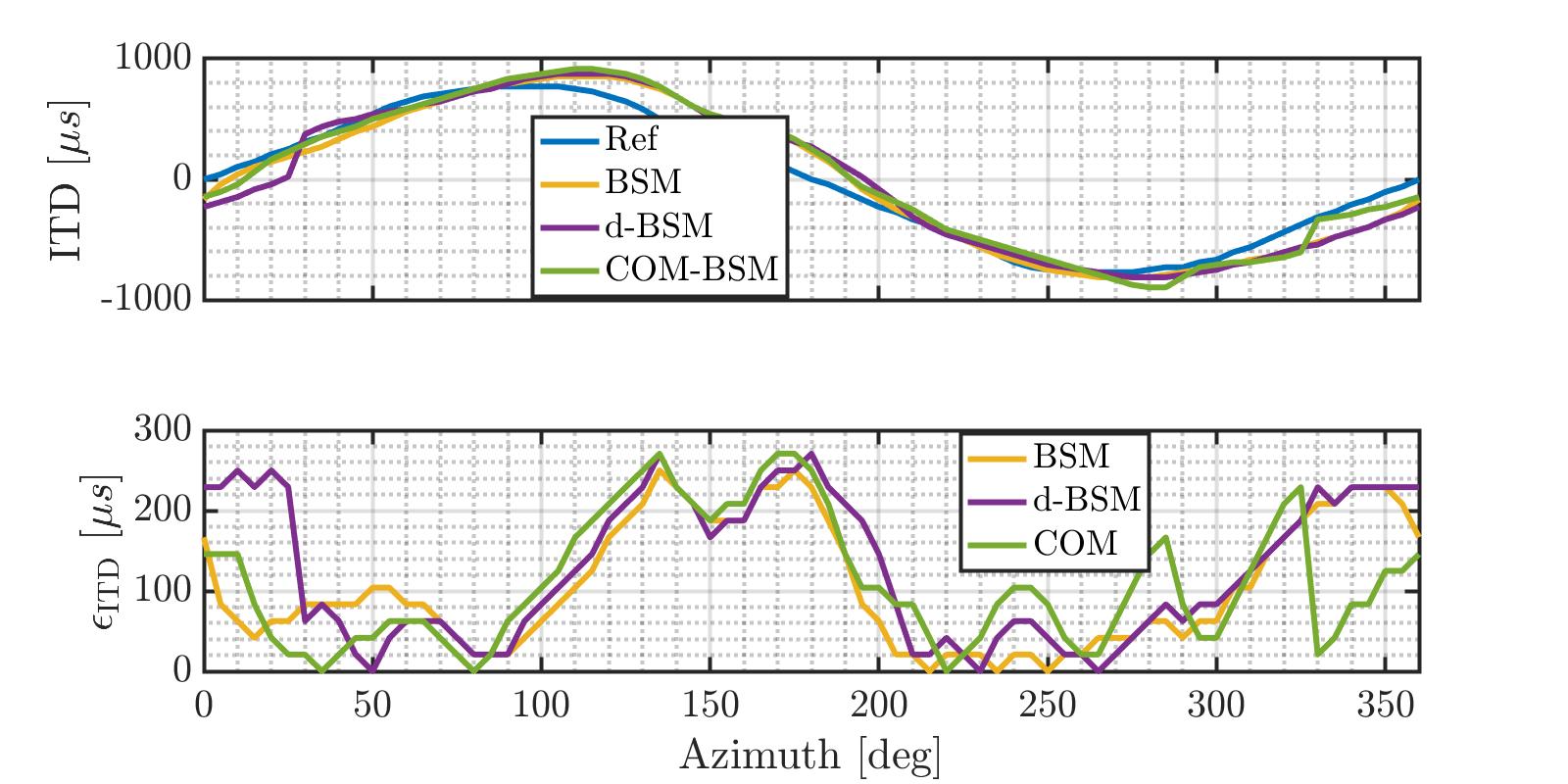}\label{fig:ITD_30_new}}
\subfigure[]{\includegraphics[width=9cm]{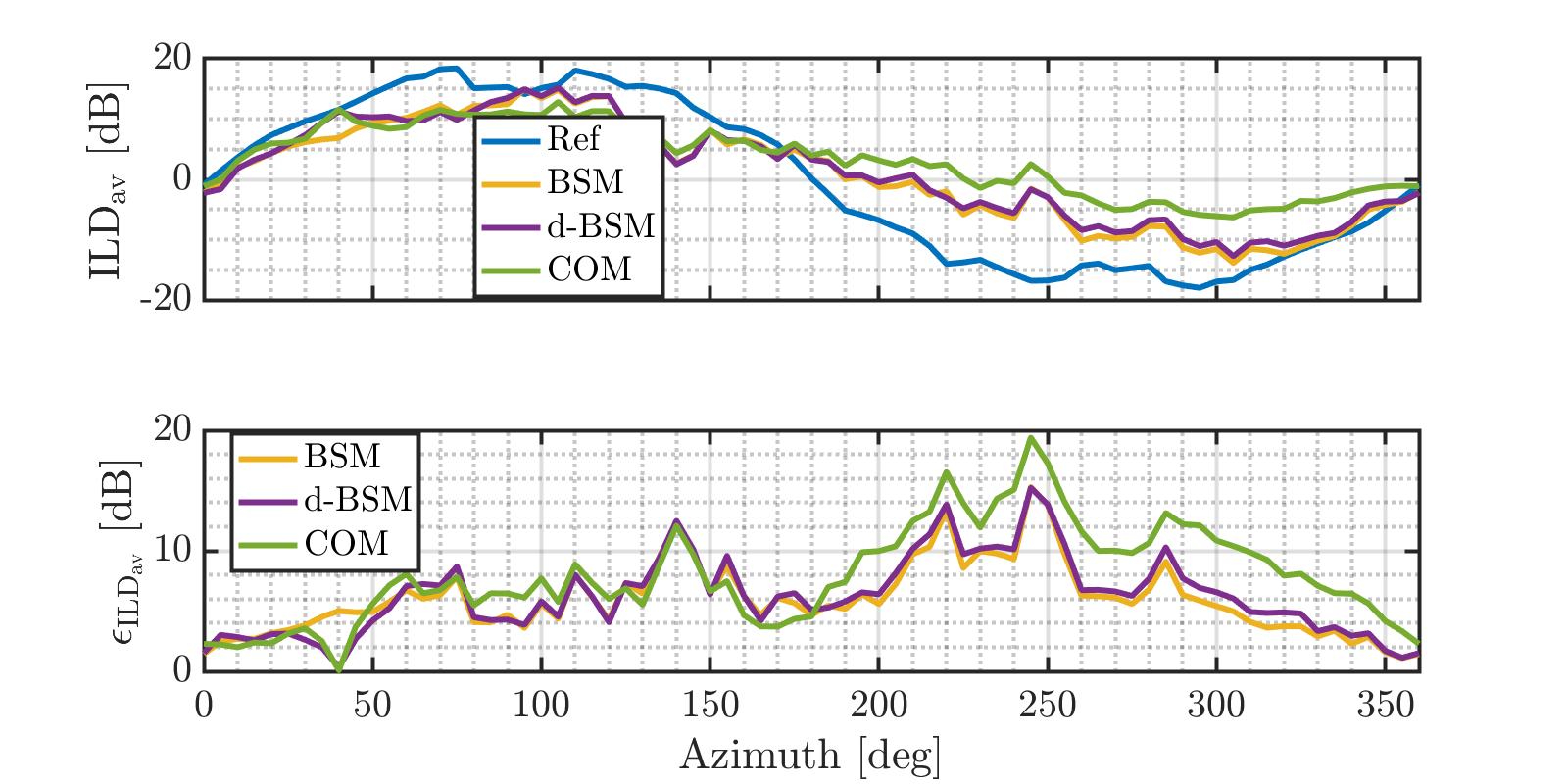}\label{fig:ILD_30_new}}
\caption{The ITD (a) and ILD (b) of scenario 1 in Table \ref{en:sims}, with the direct source positioned at a single direction of $40^\circ$. The orange line represents the BSM method, the purple line corresponds to the d-BSM approach, and the green line depicts the COM approach. ITD and ILD errors are also presented, as defined in Section \ref{sec:ITD&ILD}. $50^\circ$ head rotation was employed.}
\label{fig:dir_30}
\end{figure}

Figure \ref{fig:dir_30} illustrates the performance of ILD and ITD as detailed above 
with a $50^\circ$ rotation of the head.
Figure \ref{fig:dir_30} shows that while the methods show similar results, some important observations can be made. It can be seen that while the ILD performance of the three methods is similar in most directions, on the direct source direction, $40^\circ$, both d-BSM and COM possess a lower error than the BSM, as expected. Also note that for the COM method, there is a slightly higher error in the directions opposite from the source, e.g. around $220^\circ$,while the d-BSM performance remains consistent with the BSM method in those directions.
Furthermore, a discernible trade-off can be revealed wherein the enhancement of BSM performance in the DOA of a sound source is counterbalanced by a marginal decrement in performance across other directions. In particular, d-BSM performs better in directions away from the source than COM.

\section{Listening Experiment}
The simulation results in the previous section offer valuable insights into the objective performance of the different BSM approaches, but they do not fully capture how these differences are perceived by human listeners. This section describes a listening experiment designed to subjectively compare the quality of the BSM, d-BSM, and COM methods.

\subsection{Setup}
\label{sec:exp-setup}
The setup in this experiment is based on a simulation of a point source inside a room according to scenario 1 described in Table \ref{en:sims}. The source is positioned at an angle of ${\Omega_d}=(40^\circ,90^\circ)$ around the array, which was also simulated as described in Section \ref{sec:set-up}, i.e. a semi-circular array with 6 microphones. The room impulse response from the source to the array microphones was simulated using the image method \cite{ref19} in MATLAB \cite{ref44}. Microphone signals were generated by convolving the room impulse responses with the speech signal detailed in Table \ref{en:sims}. HRTF was also used as detailed in Section \ref{sec:set-up}, including head rotations.

\subsection{Methodology}
The binaural signals in this experiment were generated according to Section \ref{sec:methodology}.
The study employed the Multiple Stimuli with Hidden Reference and Anchor (MUSHRA) test \cite{ref30}. The test featured two MUSHRA screens: both with $50^\circ$ head rotation, one with an error of $10^\circ$ in the estimation of speaker direction, as shown in Section \ref{sec:Robus} and one without. Within each screen, the reference signal was a HOA signal of order $N = 14$, detailed in Section \ref{sec:methodology}. Each screen included the following four test signals: the hidden reference (HOA), the anchor BSM, d-BSM, and COM, implemented as detailed in Section \ref{sec:methodology}.
Ten subjects, all reported to have normal hearing, participated in the study and were presented with MUSHRA screens and signals in a randomized sequence.
They evaluated the similarity of each test signal to the reference signal based on overall quality, encompassing spatial and timbral characteristics. Scores ranged from 0 to 100, with 100 indicating that the test signal was indistinguishable from the reference.
In addition, headphone compensation filters cited in \cite{ref19} were utilized. Prior to the listening test, participants underwent two training stages: the first familiarized them with the scoring criteria, while the second introduced them to the test signals without requiring scoring. A Repeated Measures ANalysis of VAriance (RM-ANOVA) \cite{ref45} with two within-subject factors and their interaction was conducted on the scores.


\subsection{Results}

\begin{figure}[ht]
\centering
{\includegraphics[width=9cm]{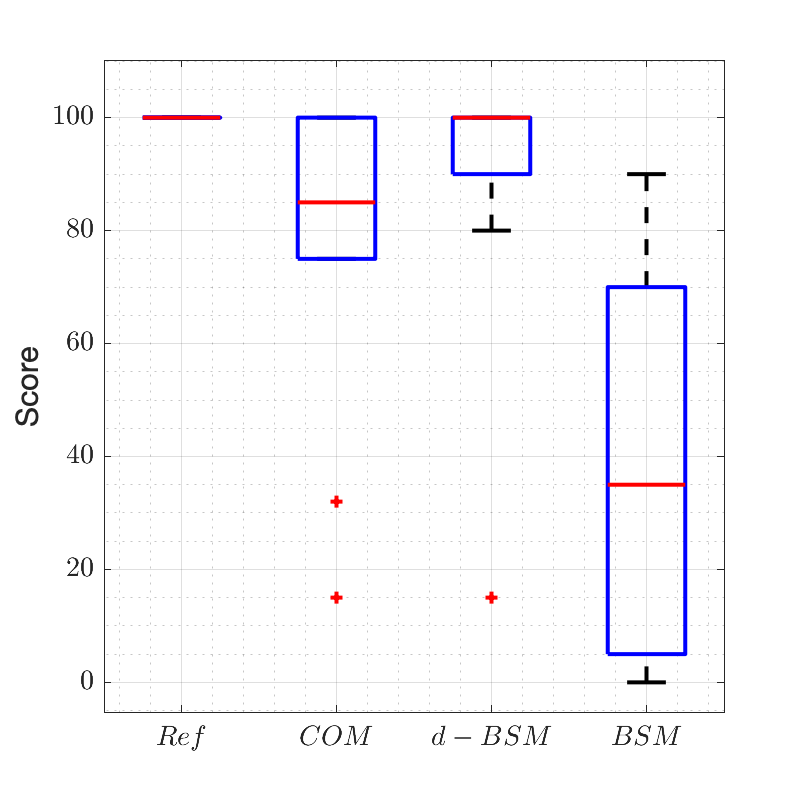}}
\caption{Similarity to the reference with respect to overall quality, displayed in a box plot, comparing four signals without estimation error: Reference, COM, d-BSM, and BSM, generated as explained in Section \ref{sec:exp-setup}. In the plot, the red line indicates the median, the box edges denote the 25th and 75th percentiles, and the whiskers represent the range of values, excluding outliers.}
\label{fig:listening1}
\end{figure} 

Figure \ref{fig:listening1} displays the results of the listening experiment conducted as described above, assuming no estimation error. The reference value consistently received 100 points with no variance, indicating it was identified correctly every time. Both the COM and d-BSM methods received higher scores than the BSM, suggesting an improvement.
The RM-ANOVA analysis indicated a significant main effect, with $F(3,27)=8.7$, $p<0.001$, and $\eta^2_p=0.4918$. Due to this significant effect, a post-hoc test with Bonferroni correction was performed for further analysis. Pairwise comparisons of the estimated marginal means showed no significant difference between the reference and the COM and d-BSM methods, with a mean difference of $11.5$ and $22.8$, respectively, and $p>0.2425$. The most significant difference was between the BSM and the reference, with $p<0.001$ and a mean difference of $61.7$. These results suggest that the adaptive methods improved the BSM performance in this case.

\begin{figure}[ht]
\centering
{\includegraphics[width=9cm]{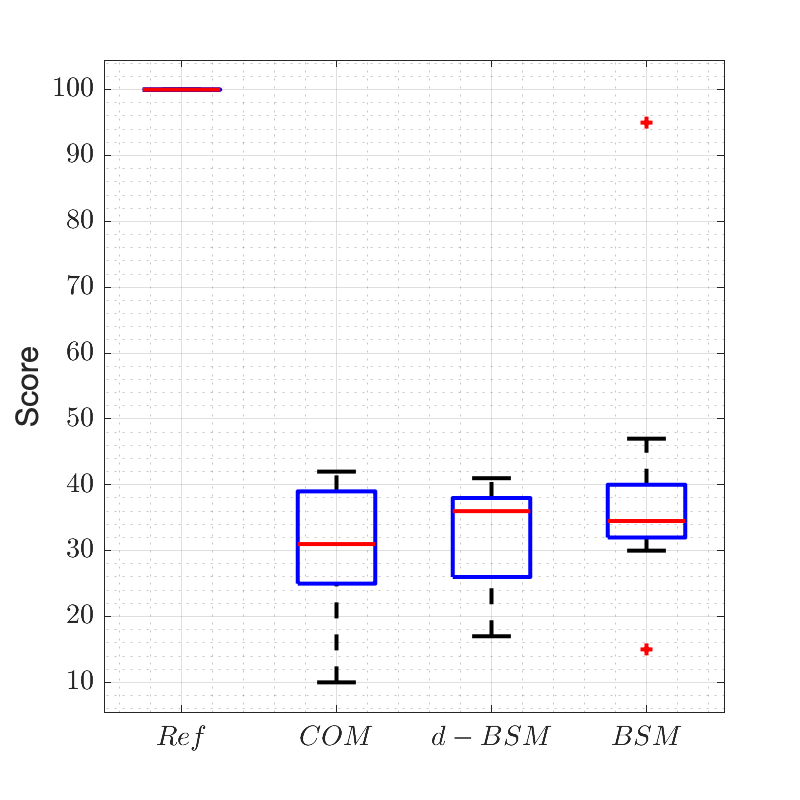}}
\caption{Similarity to the reference with respect to overall quality, displayed in a box plot, comparing four signals with $10^\circ$ of error in the estimation of speaker direction: Reference, COM, d-BSM, and BSM, generated as explained in Section \ref{sec:exp-setup}. In the plot, the red line indicates the median, the box edges denote the 25th and 75th percentiles, and the whiskers represent the range of values, excluding outliers.}
\label{fig:listening2}
\end{figure} 

Figure \ref{fig:listening2} shows the results of the listening experiment conducted as described above, assuming a $10^\circ$ error in the estimation of speaker direction. Again, the reference value consistently received a score of 100 with no variance, as expected. Both the COM, d-BSM, and BSM methods received similar scores. The RM-ANOVA analysis indicated a significant main effect, with $F(3,27)=65.44$, $p<0.001$, and $\eta^2_p=0.88$. Due to this significant effect, a post-hoc test with Bonferroni correction was performed for further analysis. Pairwise comparisons of the estimated marginal means showed no significant difference between the BSM, COM, and d-BSM methods, with a mean difference smaller than $9$ and $p=1$ for all of them. The most significant difference was between the reference and all other methods, with $p<0.001$ and a mean difference higher than $60$. These results suggest that when incorporated with estimation error, both d-BSM and COM converge to the BSM performance.

\section{Conclusions}
This paper offers several significant contributions to the field of binaural signal matching and its application with wearable microphone arrays. 
The paper presents, analyzes, and compares two approaches that incorporate signal information into the BSM method, specifically focusing on source DOA. These methods effectively address the limitations imposed by the \textcolor{\MyColor}{diffuse-field assumption of} BSM, demonstrating improved performance across source directions. This enhanced performance was also confirmed through ITD and ILD analysis.
Furthermore, the performance analysis revealed a only a minor trade-off, where enhancing BSM performance for the DOA of a sound source leads to a slight decrease in performance across other directions. One of the proposed methods showed superior performance in directions away from the source and was found to be more robust to errors in the estimation of DOA.
Finally, the paper includes a comprehensive simulation study and a listening experiment that validate the accuracy and quality of the different methods using a wearable microphone array. The results confirm the effectiveness of the proposed approaches and highlight their potential for practical applications.
Despite these advances, several limitations remain. The performance of the proposed methods can be significantly affected by errors in DOA estimation, which was evident from the robustness analysis. Additionally, the methods were primarily tested in controlled environments, and their performance in more complex real-world scenarios remains to be fully explored.
Future work could focus on improving DOA estimation accuracy to further enhance the robustness of the proposed methods. Additionally, extending the validation of these methods to diverse and dynamic real-world environments would be valuable. Investigating the integration of these methods with other spatial audio processing techniques could also offer new insights and improvements in binaural signal reproduction.

\section{Acknowledgment}
This work was partially supported by Reality Labs Research @ Meta.

\bibliographystyle{abbrv}
\renewcommand{\refname}{\normalfont\selectfont\normalsize}

\end{document}